\documentclass[fleqn,usenatbib]{mnras}

\usepackage{newtxtext,newtxmath}
\usepackage[T1]{fontenc}

\DeclareRobustCommand{\VAN}[3]{#2}
\let\VANthebibliography\thebibliography
\def\thebibliography{\DeclareRobustCommand{\VAN}[3]{##3}\VANthebibliography}

\usepackage{graphicx}	% Including figure files
\usepackage{subcaption} % Including Subcaption
\usepackage{amsmath}	% Advanced maths commands
\usepackage{soul}
\usepackage{bm}
\usepackage{tabularx}

\usepackage[encapsulated]{CJK}
\usepackage{ucs}
\usepackage[utf8]{inputenc}
\usepackage{xcolor}
\usepackage{physics}
\usepackage{tikz}

\usetikzlibrary{shapes.geometric, arrows}
\tikzstyle{startstop} = [rectangle, rounded corners, minimum width=2cm, minimum height=1cm,text centered, draw=black, fill=red!30]
\tikzstyle{midprod} = [rectangle, rounded corners, minimum width=2cm, minimum height=1cm,text centered, draw=black, fill=blue!30]
\tikzstyle{process} = [rectangle, minimum width=3cm, minimum height=1cm, text centered, draw=black, fill=orange!30]
\tikzstyle{decision} = [diamond, minimum width=3cm, minimum height=1cm, text centered, draw=black, fill=green!30]
\tikzstyle{arrow} = [thick,->,>=stealth]

\newcommand{\tctext}[1]{\begin{CJK}{UTF8}{bkai}#1\ignorespacesafterend\end{CJK}}

\setstcolor{red}
\usepackage[normalem]{ulem}
\newcommand\redout{\bgroup\markoverwith
{\textcolor{red}{\rule[.5ex]{2pt}{0.4pt}}}\ULon}

\newcommand{\sigunit}{g cm$^{-2}$}

\newcommand{\au}{\mathrm{au}}
\newcommand{\tm}{\mathrm{M}}
\newcommand{\lrbrac}[1]{\left({#1}\right)}
\newcommand{\expe}[1]{\langle {#1}\rangle}

\newcommand{\mbf}[1]{\mathbf{#1}}
\newcommand{\mrm}[1]{\mathrm{#1}}
\newcommand{\fref}[1]{Figure \ref{#1}}
\newcommand{\tref}[1]{Table \ref{#1}}
\title[Dusty spiral arms]{Evolution of dust in a protoplanetary disc driven by stellar flybys: implications for the streaming instability}

\author[Su et al.]{Wei-Shan Su (\tctext{蘇唯善}),$^{1,2}$\thanks{E-mail: wssu@asiaa.sinica.edu.tw}
Jeremy L. Smallwood,$^{3,1}$\thanks{E-mail: drjeremysmallwood@gmail.com}
Min-Kai Lin  (\tctext{林明楷}),$^{1}$ 
\newauthor
Chao-Chin Yang (\tctext{楊朝欽}),$^{4}$ 
Nicol\'{a}s Cuello$^{5}$
\\
$^{1}$Institute of Astronomy and Astrophysics, Academia Sinica, Taipei 106319, Taiwan\\
$^{2}$Graduate Institute of Astrophysics, National Taiwan University, Taipei 106216, Taiwan\\
$^{3}$Homer L. Dodge Department of Physics and Astronomy, The University of Oklahoma, Norman, OK 73019, USA \\
$^{4}$Department of Physics and Astronomy, The University of Alabama, Box~870324, Tuscaloosa, AL~35487-0324, USA \\
$^{5}$Univ. Grenoble Alpes, CNRS, IPAG, 38000 Grenoble, France
}

\date{Accepted XXX. Received YYY; in original form ZZZ}

\pubyear{2025}

\begin{document}
\label{firstpage}
\pagerange{\pageref{firstpage}--\pageref{lastpage}}
\maketitle

\begin{abstract}
     Stellar flybys are a common dynamical process in young stellar clusters and can significantly reshape protoplanetary discs. However, their impact on dust dynamics remains poorly understood, particularly in the weakly coupled regime (St$\gg$1). We present three-dimensional hydrodynamical simulations of parabolic stellar flybys—both coplanar and inclined—interacting with a gaseous and dusty protoplanetary disc. Dust species with Stokes numbers ranging from 15 to 100, corresponding to four grain sizes under a uniform initial gas surface density, are included. Perturber masses of 0.1 and 1$\mathrm{M}_{\odot}$ are considered. The induced spiral structures exhibit distinct dynamical behaviours in gas and dust: dust spirals retain a nearly constant pattern speed, while gas spirals gradually decelerate. The pitch angles of both components decrease over time, with dust evolving more rapidly. In the weakly coupled regime, gas and dust spirals are spatially offset, facilitating dust accumulation around both structures. Equal-mass flybys truncate the disc at approximately $\sim$0.55$r_{\mathrm{Hill}}$, producing tightly wound, ring-like spirals that promote dust concentration. By mapping the streaming instability growth rates in the  solid abundance–Stokes number space across three evolutionary phases, we find that a low-mass flyby suppresses dust concentration below the critical clumping threshold after periastron and maintains this suppression over time, indicating long-lasting inhibition of dust clumping. An equal-mass flyby raises local solid abundance well above the threshold, suggesting that such encounters may foster conditions favourable for dust clumping. Flyby-induced spirals play a central role in shaping dust evolution, leading to distinct spatial and temporal behaviours in weakly coupled discs. 
\end{abstract}

%%%%%%%%%%%%%%%%%%%%%%%%%%%%%%%%%%%%%%%%%%%%%%%%%%

%%%%%%%%%%%%%%%%% BODY OF PAPER %%%%%%%%%%%%%%%%%%

\section{Introduction}
The role of large-scale spiral structures in protoplanetary discs, which can stretch from tens to hundreds of au, remains a key area of investigation in planet formation. These density waves can be excited by several mechanisms, including embedded planets \citep{Kley2012, Kratter2016}, disc self-gravity \citep{Baehr2021a}, or external gravitational perturbations from stellar flybys \citep{Smallwood2023}. This study focuses on the latter mechanism, using hydrodynamical simulations to investigate the response of both gas and dust within a protoplanetary disc to a stellar encounter. Our goal is to characterize the resulting structural changes and explore their implications for the process of planet formation.

A perturbing body on a flyby or unbound trajectory is characterized by a single close approach, or periastron passage, within a distance of 1000 au from the target system \citep{Clarke1993}. The likelihood of such stellar flyby events depends strongly on the local stellar density \citep{Hillenbrand1997, Carpenter2000, Lada2003, Porras2003}. Most stars are thought to form in dense clustered environments such as the Orion Nebula Cluster, where peak stellar densities reach $\sim 10^4\,{\rm stars}\,{\rm pc}^{-3}$ \citep{Hillenbrand1998}. In such environments, stellar encounters predominantly occur within the first million years of evolution \citep{Bate2018}, a timescale that overlaps with the 1--10 Myr lifetime of gaseous protoplanetary discs \citep{Haisch2001, Hernandez2007, Hernandez2008, Mamajek2009, Ribas2015}. Based on the findings of \citet{Pfalzner2013} and \citet{Winter2018a}, the probability of a solar-type star experiencing a stellar flyby within the first million years of its evolution is estimated to be approximately 30 per cent in regions with stellar densities exceeding $\sim 100\,{\rm stars}\,{\rm pc}^{-3}$. Moreover, \citet{Pfalzner2021} demonstrated that previous estimates of close flyby frequencies in low-mass clusters were conservative, suggesting that 10 to 15 per cent of discs with radii smaller than 30 au may have been significantly altered or truncated by flyby interactions. Even in low-density environments, stellar flybys remain dynamically relevant for disc evolution. Taurus, with stellar densities of only $\sim 10$--$100\,{\rm stars}\,{\rm pc}^{-3}$, represents a prototype of isolated star formation at the lower end of the density distribution observed in nearby star-forming regions \citep{Luhman2018}. \citet{Winter2024} presented a detailed dynamical model of the Taurus region, finding that approximately one quarter of discs are truncated below 30 au by dynamical encounters over the disc lifetime, with strongly truncating encounters (ejecting $\gtrsim 10$ per cent of disc mass) occurring at a rate of $\sim 10\,{\rm Myr}^{-1}$. Consequently, flyby events represent a significant mechanism for perturbing and shaping protoplanetary discs across the full range of star-forming environments \citep{Cuello2019, Cuello2020, Jimenez-Torres2020, Menard2020}.

Several observed flyby candidates exhibit interactions with protoplanetary discs, including RW Aurigae (RW Aur; \citealt{Cabrit2006, Dai2015, Rodriguez2018}), AS 205 \citep{Kurtovic2018}, HV Tauri and Do Tauri \citep{Winter2018b}, FU Orionis (FU Ori; \citealt{Beck2012, Takami2018, Perez2020, Borchert2022a, Borchert2022b}), Z Canis Majoris (Z CMa; \citealt{Takami2018, Dong2022}), UX Tauri (UX Tau; \citealt{Menard2020}), and Sagittarius C (Sgr C; \citealt{Lu2022}). The classification of V2775 Orionis (V2775 Ori; \citealt{Zurlo2017}) and V1647 Orionis (V1647 Ori; \citealt{Principe2018}) as flyby encounters remains highly speculative. For a recent review on the role of flybys in shaping protoplanetary discs, see \cite{Cuello2023}.

As a perturber approaches periastron passage, tidal effects can induce the formation of spirals and potentially lead to disc fragmentation \citep{Ostriker1994, Pfalzner2003, Shen2010, Thies2010, Smallwood2023}. External unbound companions excite spiral density waves at Lindblad and corotation resonances \citep[e.g.,][]{Lin1993}. If the unbound companion is a stellar-mass object, it exerts a strong tidal force, with its Roche lobe potentially extending beyond the location of most of these resonances. Additionally, flyby encounters can warp the primary disc across a range of perturber inclinations and periastron distances \citep{Clarke1993, Ostriker1994, Terquem1996, Bhandare2016, Xiang-Gruess2016,Nealon2020}.

Numerical simulations have provided insights into the effects of flyby encounters on protoplanetary discs. \citet{Winter2018b} demonstrated that a flyby in a parabolic orbit induces the longest interaction time and the strongest tidal effects among unbound perturbations. \citet{Cuello2019} showed that flybys can generate two-arm grand-design spirals with varying pitch angles, where dusty spirals appear sharper than gaseous spirals. \citet{Smallwood2023} investigated the evolution of the pattern speed and pitch angle of spirals induced by a low-mass flyby, finding that the initial pattern speed closely follows the angular velocity of the flyby relative to the central star. Over time, the pitch angle of the spirals decreases, leading to their gradual winding, which differs from spirals produced from bound companions and gravitational instability. Beyond spiral formation, flyby encounters can significantly reshape discs through gravitational interactions. In the case of a stellar-mass perturber, the encounter can be destructive, truncating the disc and altering its overall structure \citep{Cuello2019, Smallwood2023}, which may impact the planet formation process.

During the process of planet formation micron-sized dust aggregates can collide and adhere to one another, forming larger structures. However, when millimeter-sized grains collide, the outcomes are often less constructive, resulting in bouncing or fragmentation \citep{Blum2018}. This presents a significant challenge for the growth of dust aggregates into kilometer-sized bodies, known as planetesimals, which are crucial building blocks for planets. One potential mechanism to overcome this barrier is streaming instability, which arises from the differential velocities between the gaseous and dusty components of the protoplanetary disc \citep{Johasen2014}.

Recently, \cite{Prasad2025} conducted hydrodynamic simulations demonstrating that stellar flybys can induce long-lived substructures that act as dust traps, locally enhancing the dust-to-gas ratio to values sufficient to trigger the streaming instability. In this work, we investigate the effects of a flyby encounter on a protoplanetary disc, with a particular emphasis on its implications for planetesimal formation via the streaming instability. Unlike \cite{Prasad2025}, we focus our attention on the weakly coupled regime between gas and dust, quantify SI growth rates and also consider inclined flyby encounters. Through this exploration, we aim to uncover the relationship between external gravitational perturbations and the clumping of dust grains, a crucial step in the planet formation process. Building on the foundational work of \cite{Smallwood2023}, we extend their study by incorporating dust dynamics into our simulations. This allows us to assess and compare the morphological evolution of both the gaseous and dusty components of the disc following a flyby encounter, offering new insights into how such interactions influence disc structure and the formation of planetesimals. The outline of the paper is as follows. In Section~\ref{sec::methods}, we detail the hydrodynamical simulation setup and analysis techniques. Next, the results of the hydrodynamical simulations are presented in Section~\ref{chap:Result}. In Section~\ref{schap:ApplicationSI}, we discuss the results of the simulations in the context of planetesimal formation via the streaming instability. We present the caveats of our study in Section~\ref{sec::Caveats}, and give our conclusions in Section~\ref{sec::conclusions}.
 \begin{figure*}
    \centering
    \includegraphics[width=2\columnwidth]{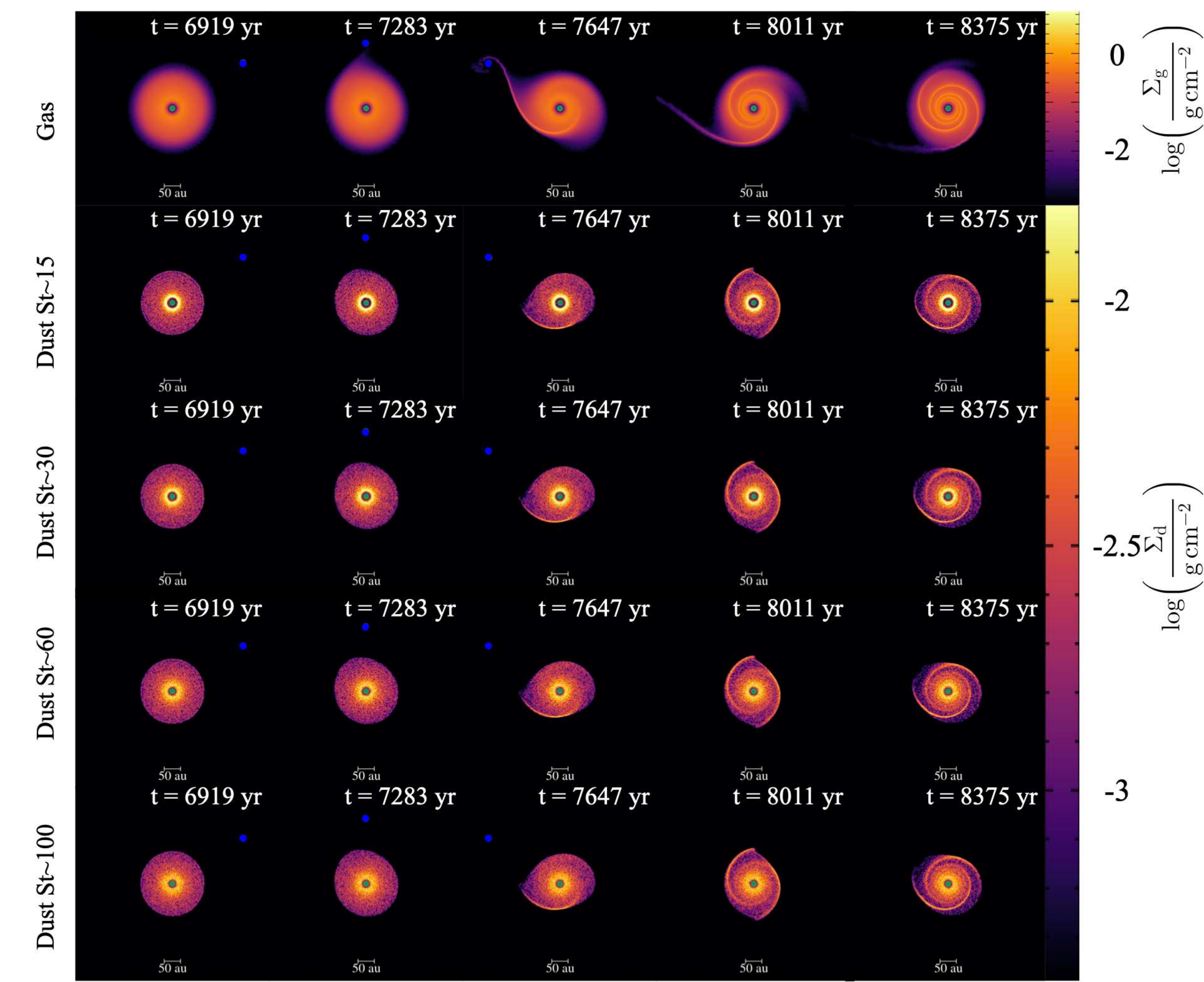}
    \caption{Evolution of the column density of gaseous and dusty components, with various initial Stokes numbers, during a low-mass coplanar flyby. We display only the dynamics of the gas disc from simulation ID st15co, as the gas disc behavior is nearly identical across simulations of the same flyby passage. The remaining rows show dust discs with various initial Stokes numbers. Blue dots mark the location of the perturbing companion, and green denotes the central star.}
    \label{fig:Low-mass_Splash_movie}
\end{figure*}
\section{Methods}% Describe simulation setup
\label{sec::methods}

In order to investigate how unbound perturbers influence protoplanetary discs with gas and dust, we conduct 3-dimensional smoothed particle hydrodynamical (SPH) simulations using {\sc phantom} \citep{Price2018}. Here, the term {\it dust} follows the convention used in SPH and hydrodynamic studies \cite[e.g.,][]{Laibe2012a, Price2015a, Dipierro2015}, referring to the solid component of the disc regardless of size, as long as it is dynamically coupled to the gas via the drag force between gas and dust.  This SPH code has well tested for simulating perturbers on flyby orbits \cite[e.g.,][]{Cuello2019,Cuello2020,Borchert2022a,Borchert2022b,Smallwood2023,Smallwood2024}. We use a similar simulation setup as \cite{Smallwood2023}, but consider both gas and dust dynamics.

The degree of aerodynamic coupling between dust particles and gas is quantified by the dimensionless Stokes number (St). This parameter is defined as the ratio of the particle stopping time to a characteristic dynamical timescale of the flow. The stopping time represents the exponential decay timescale of the dust-gas relative motion due to drag forces. In protoplanetary discs, the relevant dynamical timescale is typically the inverse of the local Keplerian frequency. For particles in the Epstein drag regime \citep{Weidenschilling1977}, the Stokes number is given by:
\begin{equation}
\label{eq:Stokesnumber}
     \mathrm{St} = \frac{\pi}{2} \frac{a\rho_{\rm \bullet}}{\Sigma_{\rm g}},
\end{equation}
where $a$ is the dust size, $\Sigma_{\rm g}$ is the gas column density, and $\rho_{\rm \bullet}$ is the dust intrinsic density, and is set to $3\, \rm g\,cm^{-3}$. Each simulation adopts a single dust size, with $a = 1,\,2,\,4,$ and $7\,\mathrm{cm}$ representing the four grain-size setups explored together with various flyby parameters described in Section \ref{chap:flyby-setup}. In {\sc phantom}, the representative initial Stokes number is not derived from particle statistics but analytically estimated from the disc setup. Specifically, {\sc phantom} evaluates the approximated initial Stokes number at the midpoint radius of each disc, following the Eq.~\ref{eq:Stokesnumber}. The resulting approximate values are ${\mathrm{St}}_{\mathrm{init.}} \sim 15,\,30,\,60,$ and $100$, for grain sizes $a = 1,\,2,\,4,$ and $7\,\mathrm{cm}$, respectively. The dust size is kept constant throughout each simulation, without grain growth or fragmentation. Solid particles with $\rm St \ll 1$ are strongly coupled to the gas, while the ones with $\rm St \gg 1$ are weakly coupled.  In the current version of {\sc phantom}, the modeling of dust-gas mixtures depends on the St of the dust grains, using two different techniques.  For grains with $\rm St > 1$, {\sc phantom} simulates two distinct particle types (two-fluid method), as described in \cite{Laibe2012a,Laibe2012b}. For smaller grains with $\rm St < 1$, a single particle type is used to represent a mixture of both dust and gas (one-fluid method), following \cite{Price2015a}. In the two-fluid approach, dust and gas are treated separately, with a drag term and explicit time-stepping, while in the one-fluid method, dust is treated as part of the overall mixture, with evolution equations that account for the dust fraction. In our study, we focus on an unbound companion perturbing a dusty disc with $\rm St > 1$, therefore, we employ the two-fluid algorithm. This method includes drag heating but ignores thermal coupling between the gas and dust \cite[see][]{Laibe2012a}. The adopted Stokes numbers are chosen to probe the weakly coupled regime. For moderate Stokes numbers, around unity ($\mathrm{St} \sim 1$), the two-fluid method becomes numerically inefficient. Although these values are larger than those typically expected from dust-evolution models \cite[e.g.][]{Birnstiel+2012}, our goal here is to investigate the limiting behaviour of weakly coupled dust towards ${\rm St\sim 1}$.

\subsection{Flyby set-up}\label{chap:flyby-setup}
We model an unbound perturber on a parabolic orbit ($e = 1$) around a primary star with a protoplanetary disc. We use a Cartesian coordinate system ($x$, $y$, $z$), where the $y$-axis aligns with the direction of the argument of pericenter, and the $z$-axis corresponds to the direction of the total angular momentum vector of the circumprimary disc. 
The distance between the perturber on a parabolic orbit and the central star, $r_2$, in our model is expressed as
\begin{equation}
\label{eq:flybyorbit}
     r_2 = \frac{2r_{\rm p}}{1 + \sin\theta},
\end{equation}
where $r_{\rm p}$ is the periastron distance, which occurs at $\theta = +\pi / 2$. The probability of an encounter with a periastron distance of $100$–$1000\, \rm au$ is approximately $30$ per cent, but it rapidly decreases for increasing stellar cluster age for leaky cluster environments \cite[e.g.,][]{Pfalzner2013}. Based on this, we select a periastron distance of $r_{\rm p} = 200\, \rm au$ for all simulations. To minimize any impact on the initial conditions of the protoplanetary disc, the perturber is initially placed at a distance ten times the periastron distance.

The angular speed of the flyby companion as a function of $r_2$ is given by
\begin{align}
    \omega(r_2) &=
    \sqrt{\frac{2G(M_1 + M_2)}{r_{\rm p}^3}}
    \left(\frac{r_{\rm p}}{r_2}\right)^2\\&=
    \left(0\fdg18\,\textrm{yr}^{-1}\right)
    \left(\frac{M_1 + M_2}{M_{\sun}}\right)^{1/2}
    \left(\frac{r_{\rm p}}{200\,\textrm{au}}\right)^{-3/2}
    \left(\frac{r_{\rm p}}{r_2}\right)^2,
    \label{eq::omega}
\end{align}
where $G$ is the gravitational constant, $M_1$ is the mass of the primary star, and $M_2$ is the mass of the perturber. The relationship between the  radial distance, $r_2$, and time $t$ is given by
\begin{equation}
    \left(\frac{r_2}{r_{\rm p}} + 2 \right) \sqrt{\frac{r_2}{r_{\rm p}} - 1} = \frac{3|t - t_\mathrm{p}|}{2} \sqrt{\frac{2G (M_1 + M_2)}{r_{\rm p}^3}},
    \label{eq::rp}
\end{equation}
where $r_2 = r_\mathrm{p}$ when $t = t_\mathrm{p}$. Parabolic encounters induce the strongest tidal effects and the longest interaction time with the disc compared to hyperbolic encounters \citep{Winter2018b, Cuello2019}. We consider a central star with $M_1 = 1\, \rm M_\odot$ and  perturbers with masses of  $M_2 = 0.1\, \rm M_\odot$, and $M_2 = 1\, \rm M_\odot$. For the low-mass perturber, the perihelion time is $t_{\rm p} = 7286$ years, while for the equal-mass perturber, it is $t_{\rm p} = 5406$ years.
 
We conduct a suite of simulations of flybys on both prograde coplanar and inclined trajectories, with the system's centre of mass positioned at the origin and the tilt of the flyby orbit measured relative to the $z$-axis. In the coplanar case, the perturber moves within the $x$–$y$ plane (its angular momentum vector aligned with the $z$-axis), entering from the fourth quadrant and exiting the third quadrant. We also explore prograde inclined trajectories, tilted by $45^\circ$ and rotated clockwise about the $y$-axis. In both the coplanar and inclined cases, the 3D position of the pericentre remains unchanged. The periastron occurs at $x=0$, $y>0$, and $z=0$. The accretion radius for the central star is set to $10\, \rm au$, matching the initial inner edge of the disc, while for the perturber, it is set to $1\, \rm au$. The accretion radius acts as a hard boundary, where any particles crossing it would be removed, and its mass is added to the corresponding object (the star or the perturber). The suite of simulations are shown in Table \ref{table:setuptable}.

\subsection{Circumprimary disc set-up}\label{chap:disc-setup}
We model a protoplanetary disc with gas and dust components. The disc is initially non-warped and lies in the $x-y$ plane. The disc initially contains 500,000 gaseous Lagrangian particles with a total mass of $0.001\ \tm_{\odot}$ and 50,000 dust Lagrangian particles with a total mass of $10^{-5}\ \tm_\odot$. The initial dust-to-gas ratio is, therefore $\varepsilon = 0.01$. Given the low-mass disc, the self-gravity of particles is ignored. The choice of the disc mass ensures that the periastron distance remains consistent with the two-body estimate (see Eqs.~\ref{eq:flybyorbit}--\ref{eq::rp}); a more massive disc would gravitationally perturb the orbit and shift the periastron inward, resulting in a smaller $r_p$ than the two-body prediction. Hence we limit the disc mass to less than 0.01\% of the primary’s mass. Increasing the disc mass, especially the gas component, would naturally raise the overall gas surface density and thus decrease the Stokes number according to Eq.~\ref{eq:Stokesnumber}.  Based on previous works, the gas spiral pattern remains largely similar even for more massive discs \cite[e.g.,][]{Cuello2019,Cuello2020,Prasad2025}.

The disc has an inner radius of $r_\mrm{in} = 10\ \au$ and an outer radius of $r_\mrm{out} = 100\ \au$. The column density of the disc is given by
\begin{equation}\label{eq:SurfaceDensityProfile}
    \Sigma_{\rm g/d} (r) = \Sigma_{\rm 0,g/d}  \lrbrac{\frac{r}{r_\mrm{out}}}^{-p}\lrbrac{1-\sqrt{\frac{r_\mrm{in}}{r}}},
\end{equation}
where $\Sigma_{\rm 0,g} = 2.21\times 10^{-1}$\sigunit for gaseous component and $\Sigma_{\rm 0,d} = 2.21\times 10^{-3}$\sigunit for dust component. We set $p = +3/2$ as for the minimum mass solar nebula \citep{Hayashi1981}. From the initial gas surface density profile, the Stokes number ranges roughly from 6 to 329 across the disc, and the range becomes even broader as the system evolves. We select a locally isothermal equation of state \cite[e.g.,][]{Lodato2007b}, such that the disc thickness is scaled with radius as
 \begin{equation}
    H = \frac{c_{\rm s}}{\Omega} \propto r^{3/2-q}, 
\end{equation}
where $\Omega = \sqrt{GM_1/r^3}$ and $c_{\rm s}$ is proportional to $r^{-q}$. Dust discs share the same initial scale height as gas discs.  We use $q = +1/2$ so that the initial disc aspect ratio  is constant at $H/r = 0.05$.

We include the \cite{Shakura1973} $\alpha_{\mrm{SS}}$ viscosity prescription that is given by
\begin{equation}
    \nu = \alpha_{\mrm{SS}} c_s H
\end{equation}
where $\nu$ represents the kinematic viscosity. For this purpose, we employ the prescription provided in \cite{Lodato2010}, which is given as
\begin{equation}
    \alpha_{\mrm{SS}} \approx \frac{\alpha_{\mrm{AV}}}{10}\frac{\expe{h}}{H}
\end{equation}
where $\expe{h}$ is the mean smoothing length on particles in a cylindrical ring at a given radius \citep{Price2018}, and $\alpha_{\mrm{AV}}$ is the artificial viscosity \citep{Price2018}. We set $\alpha_{\mrm{SS}} = 0.005$ for the desired viscosity, which sets the artificial viscosity to $\alpha_{\rm AV} = 0.176$ .  It is important to note that an $\alpha_{\rm AV}$ value of 0.1 represents the lower threshold, below which physical viscosity is not adequately resolved in SPH. Viscosities below this limit lead to disc spreading that is independent of the $\alpha_{\rm AV}$ value \cite[see][for details]{Meru2012}. Additionally, in SPH, we incorporate a parameter, $\beta_{\rm AV}$, which introduces a non-linear term initially designed to prevent particle-particle penetration in high Mach number shocks \cite[e.g.,][]{Monaghan1989}. Typically, $\beta_{\rm AV}$ is set to 2.0.  The smoothing length is defined as a function of the particle number density $n$ by
\begin{equation}
    h = h_{\mrm{fact}}n^{-1/3}
\end{equation}
where $h_{\mrm{fact}}$ is a proportionality factor that relates the smoothing length to the mean local particle spacing. The disc is resolved with an initial shell-averaged smoothing length per scale height of $\expe{h}/H \approx 0.2838$ in our simulations.

\begin{table}
	\centering
	\caption{A summary of the SPH simulation suite. The columns, from left to right, list the simulation ID, flyby orbital inclination $i_2$, flyby mass $M_2$, dust grain size $a$ and the averaged initial Stokes number $\langle \mathrm{St} \rangle_{\mathrm{init.}}$.}
	\label{table:setuptable}
	\begin{tabularx}{\columnwidth}{p{0.25\columnwidth} X X X X} % four columns, alignment for each
        \hline
        Simulation ID & $i_2/^{\circ}$ & $M_2/\tm_{\odot}$ & $a$/cm & $ \mathrm{St} _{\mathrm{init.}}\!\sim$\\ 
        \hline
        st15co & 0 & 0.1 & 1 & 15 \\
        st30co & 0 & 0.1 & 2 & 30  \\
        st60co & 0 & 0.1 & 4 & 60  \\
        st100co & 0 & 0.1 & 7 & 100  \\
        st15in & 45 & 0.1 & 1 & 15  \\
        st30in & 45 & 0.1 & 2 & 30  \\
        st60in & 45 & 0.1 & 4 & 60  \\
        st100in & 45 & 0.1 & 7 & 100  \\
        st15co-eqmass & 0 & 1.0 & 1 & 15  \\
        st30co-eqmass & 0 & 1.0 & 2 & 30  \\
        st60co-eqmass & 0 & 1.0 & 4 & 60  \\
        st100co-eqmass & 0 & 1.0 & 7 & 100  \\
        st60in-eqmass & 45 & 1.0 & 4 & 60  \\
		\hline
	\end{tabularx}
\end{table}
\subsection{Analysis}\label{chap:analysis-method}
\begin{figure*}
    \begin{subfigure}[b]{0.45\textwidth}
    \centering
    \includegraphics[width=1\columnwidth]{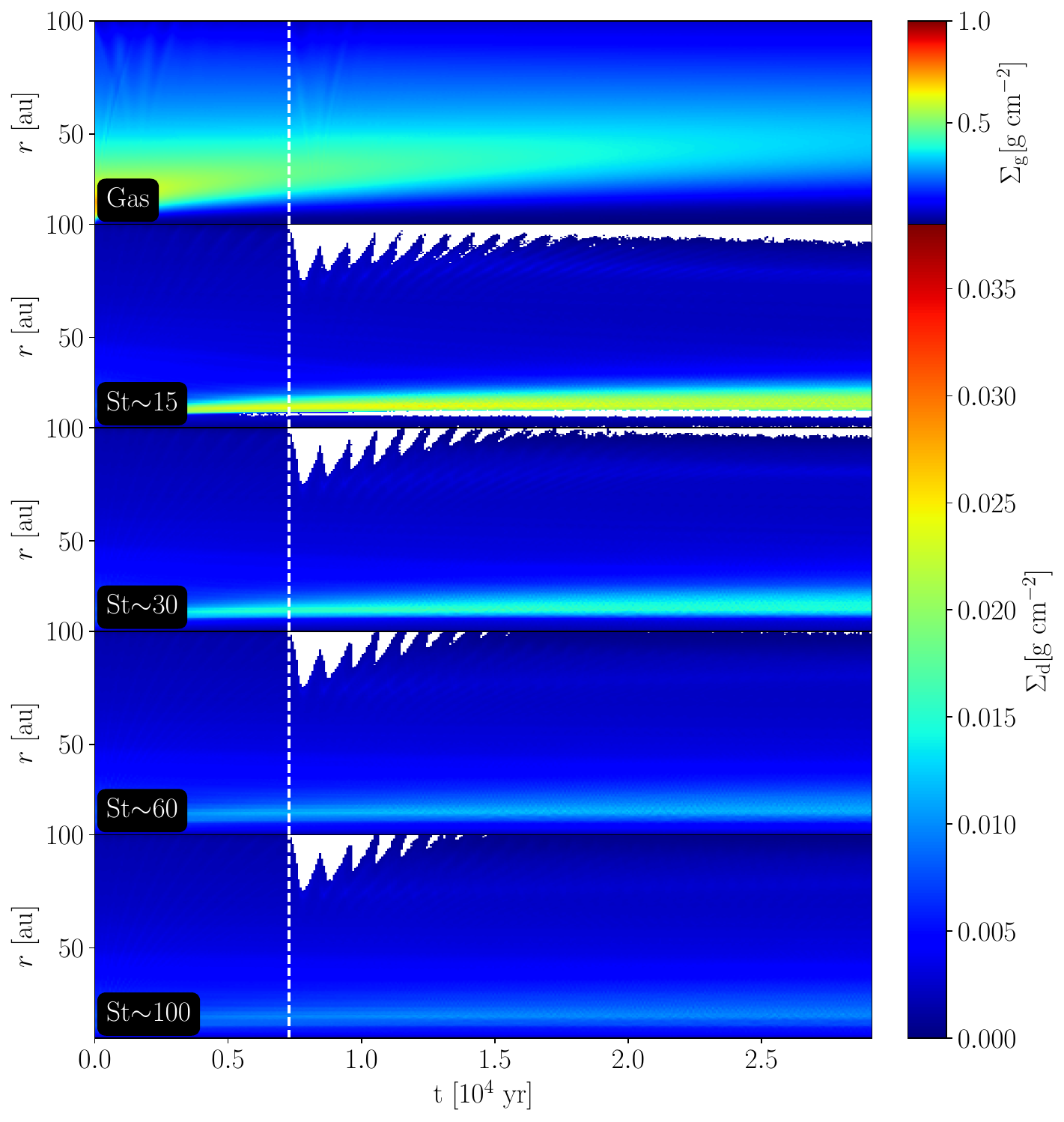}
    \caption{Azimuthally-averaged column density profile}
    \label{fig:AzimuthallyAveragedSDP}
    \end{subfigure}
    \begin{subfigure}[b]{0.45\textwidth}
    \centering
    \includegraphics[width=1\columnwidth]{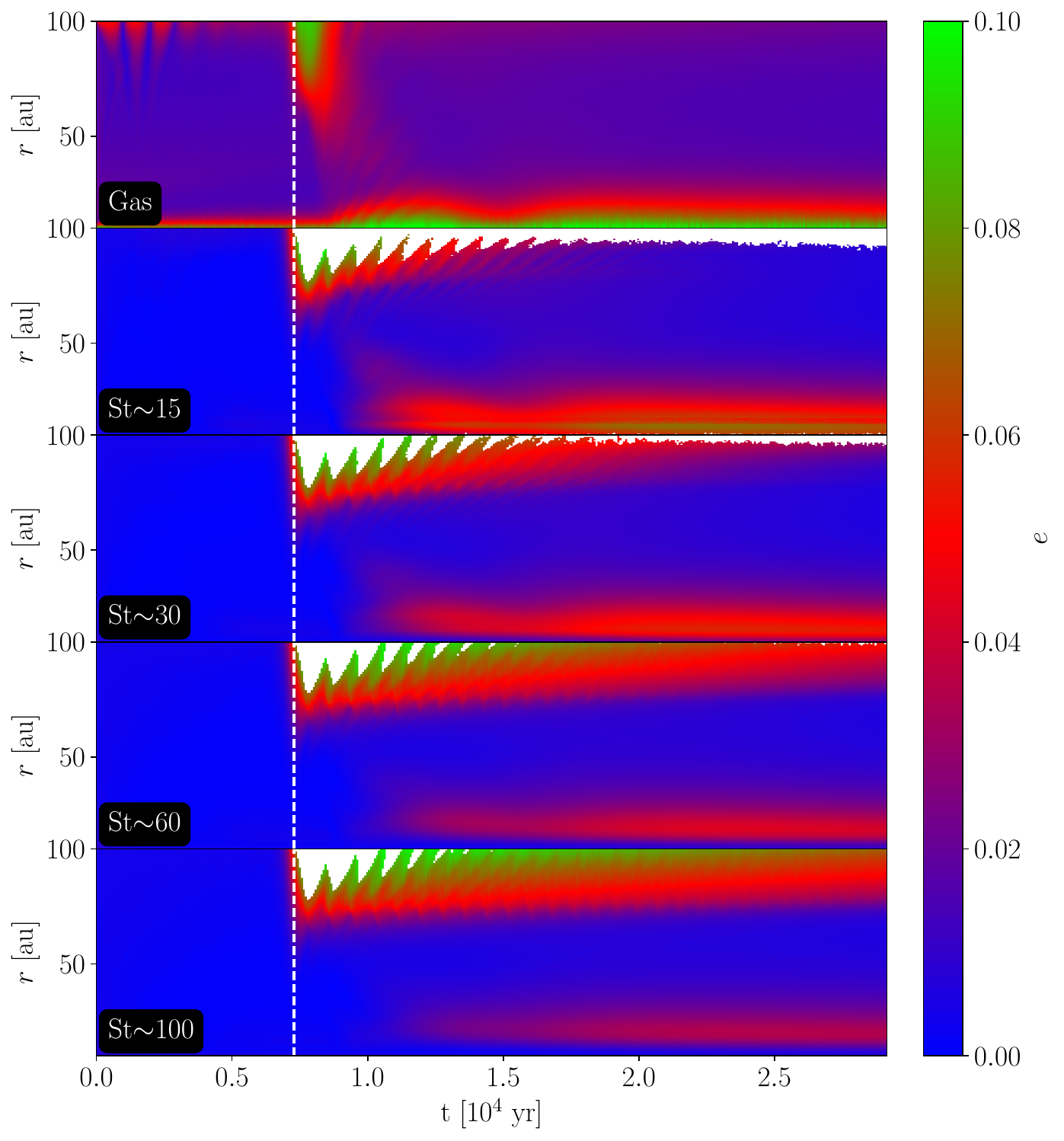}
    \caption{Azimuthally-averaged eccentricity profile}
    \label{fig:AzimuthallyAveragedECC}
    \end{subfigure}
    \caption{Evolution of the azimuthally-averaged gas and dust column densities ($\Sigma_{\rm g/d}$, left) and eccentricity ($e$, right) profiles in a radial-time grid during a coplanar low-mass flyby. White dashed lines mark the time of flyby perihelion. White regions indicate areas of zero density at the corresponding radius. The first panel in both profiles shows the gaseous disc, while subsequent panels depict dust discs with varying initial Stokes numbers. }
    \label{fig:AzimuthallyAveraged}
\end{figure*}

To analyse the dynamical features of the disc, such as spiral structures, we define a polar grid in coordinates $(r,\phi)$. The grid spans the intervals $10\, \rm au < r < 175\, \rm au$ and $0 < \phi < 2\pi$, resolved by $351 \times 351$ cells. Physical quantities are mapped from the {\sc{phantom}} SPH particle data to the grid points using kernel interpolation. For face-on dynamical studies, we compute vertically averaged quantities by integrating along the z-axis, following \cite{Price2007}.

To estimate midplane properties relevant to the streaming instability (e.g., density $\rho$, and velocity components $v_r$, $v_\phi$) \citep{Youdin2005}, we first locate the disc's effective midplane. We approximate this as the surface of maximum density, $z_{\rm max} (r,\phi)$. To determine this surface, we first construct a coarse cylindrical grid ($351\times 16\times 100$ cells in $r,\phi,z$) over the domain $z \in [-28\,\mathrm{au},28\,\mathrm{au}]$. In each $(r_i,\phi_i)$ column of this grid, we find the vertical position $z_{\rm max}$ corresponding to the maximum density. We then use linear interpolation to create a continuous $z_{\rm max} (r,\phi)$ surface. The final midplane quantities are obtained by performing a 3D kernel interpolation of the SPH data at these $(r,\phi,z_{\rm max})$ coordinates. The specific implementation of our kernel interpolation is detailed in Appendix \ref{app:KernelInterpolation}.             
\subsubsection{Pattern speed evaluation}
After extracting the information from the simulation data by kernel interpolation, we evaluate the pattern speed of spiral induced by flyby-encountering, which characterizes the rotational speed of the spiral pattern within the disc. To start with, azimuthal cuts of both gaseous and dusty column density at a specific radius $r_0$ are extracted. Here, we chose  $r_0 = 75$ au for analysis, since the dust spiral only appears at radii greater than 50 au. Thus, we selected the midpoint between 50 au and the initial outer radius. Next, we trace the spiral pattern in the azimuthal column density cuts by locating the local maxima \cite[similar to Fig.~5 in ][]{Smallwood2023}. Finally, we take the derivative of the tracings to obtain the pattern speed, $\omega$, using
\begin{equation}
    \omega = \frac{\dd}{\dd t}\left.\phi (t,r) \right|_{r=r_0},
\end{equation}
where $\phi(t)$ is the azimuth of the spiral as a function of time.

\subsubsection{Pitch angle evaluation}
The analysis conducted by \cite{Smallwood2023} showed that the flyby-triggered spiral in the outer disc is morphologically similar to a logarithmic spiral. Therefore, we developed a spiral detection technique to evaluate the pitch angle by fitting the detected spiral to the logarithmic spiral formula. The full working process of the spiral detection method is given in Appendix \ref{app:SpiralDetection}. In practice, a logarithmic spiral is expressed as a curve in the $(r, \phi)$ space by
\begin{equation}
    r(\phi) = ae^{k\phi}
\end{equation}
where $k, a$ is the parameter of  logarithmic spiral, with the conditions $a>0$ and $k\neq 0$. The pitch angle $\beta$ of  logarithmic spirals can be found in terms of $k$ as
\begin{equation}
    \beta = \tan^{-1} |k|.
\end{equation}
            
The permitted duration for spiral detection is shorter than that for the pattern speed tracing, as the spiral gradually deviates from a logarithmic shape and wrap into a tightly wound spiral at later times, exhibiting a tiny pitch angle. Consequently, distinguishing between the two spiral arms from a set of detected points becomes challenging. In our analysis, we terminate the spiral detection at an earlier time compared to the pattern speed tracing.

\begin{figure}
    \centering
    \includegraphics[width=0.98\columnwidth]{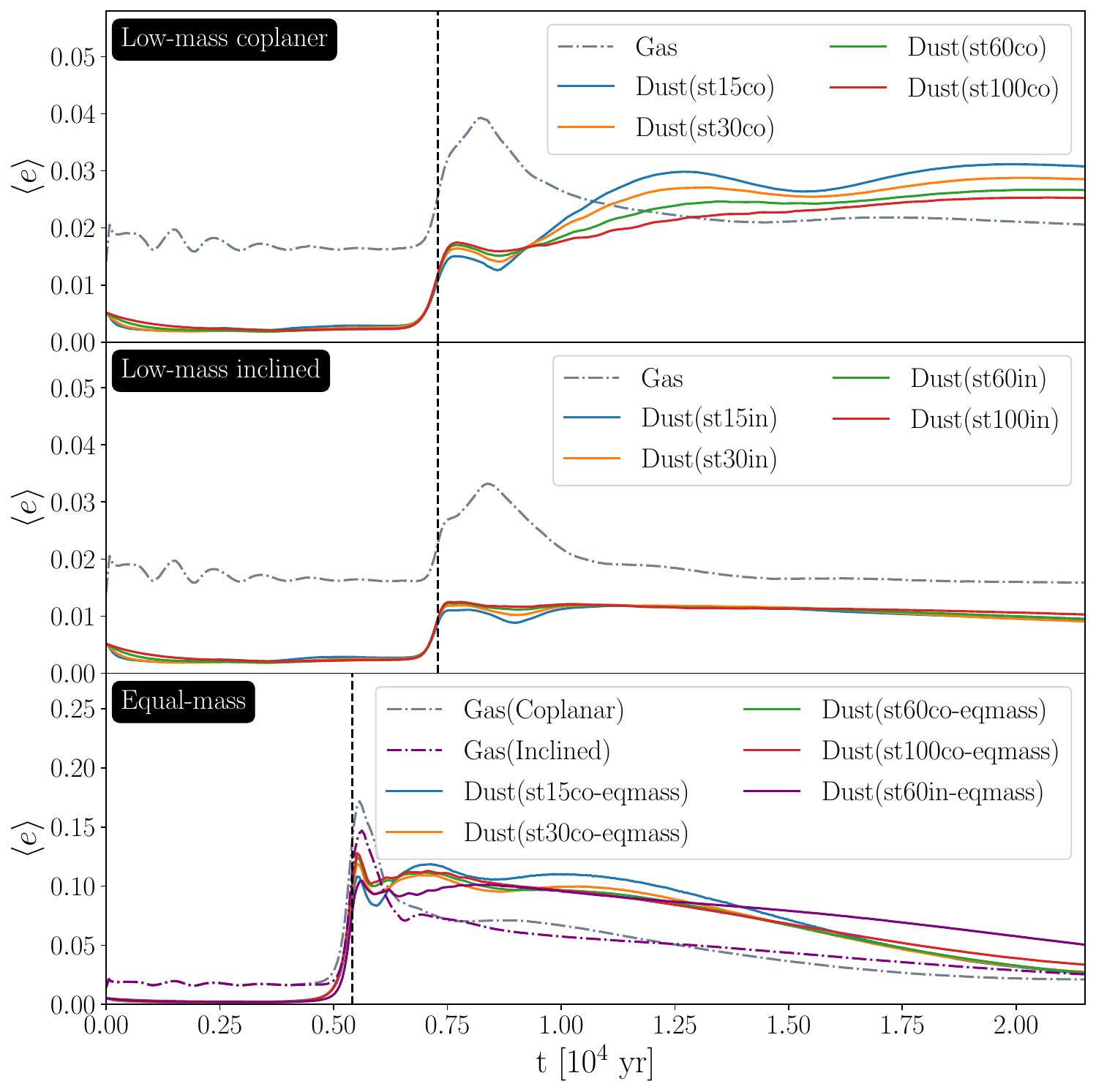}
    \caption{The evolution of the density-weighted eccentricity of the disc is shown. The vertical black line represents the time that flyby hit the perihelion for each simulation. The other non-vertical dashed lines represent the evolution of eccentricity for the gaseous disc, while the solid lines depict that for the dusty disc. Three panels show the evolution of averaged eccentricity weighted by column density encountered by low-mass coplanar, low-mass inclined and equal-mass flyby, respectively. Due to the similarity of gas disc evolution among the simulations with same central star and flyby passage, we only show the evolution from the simulations with St$\sim 15$ for each flyby setup.}
    \label{fig:WeightAveragedECC}
\end{figure}
\subsubsection{Streaming instability and critical solid abundance}\label{sschap:MethodDustGrowth}
In addition to tracing their morphology, we use the detected spiral features to estimate the growth rate of the streaming instability within the arms. For this, we adopt the method of \cite{Chen2018}, who developed a method to estimate the growth rate by transforming the problem into a eigenvalue problem, which is given as
\begin{equation}\label{eq:SIgrowthep}
    \mathbf{M\xi}=\sigma\mathbf{\xi},
\end{equation}
where $\mathbf{M}$ is the $8\times8$ matrix representation of the perturbation equations in \cite{Chen2018}, $\mathbf{\xi} = (\delta\rho_{\mathrm{d}},\delta \mathbf{v}_{\mathrm{d}},\delta_{\mathrm{g}},\delta \mathbf{v}_{\mathrm{g}})^{\mathrm{T}}$ is an eigen-vector for the streaming instability, and $\sigma$ is the corresponding eigenvalue,  with the growth rate defined as $s = \mathrm{Re}(\sigma)$. We solve Equation \ref{eq:SIgrowthep} with both diffusion and physical viscosity set to zero, thereby obtaining an idealized upper bound on the streaming instability growth rate. In practice, for each point on the polar grid, we first extract the midplane density and velocity of both gas and dust at that location. We then solve Equation \ref{eq:SIgrowthep} to obtain the growth rate and the corresponding eigenmodes at each pair of wave number $(K_x, K_z)$, where $K_{x,z} = k_{x,z}H_{\mathrm{g}}$ is the dimensionless perturbation wave number. Finally, we select the eigenmode associated with the maximum growth rate to represent the “dominant mode” at that location. We note that the streaming instability is an axisymmetric instability, and the assumption of the axisymmetry may break down in a flyby–perturbed disc. 
            
Furthermore, the growth rate of the linear modes of the streaming instability may not accurately predict strong dust clumping and planetesimal formation, and it appears that the solid abundance $Z$ is a better indicator \citep{Yang2018, Li2021}. Therefore, we adopt the latest measurement of the critical $Z$ for strong clumping as a function of St, proposed in \citet{Lim2025}. Formally, $Z$ is defined as the ratio of the column densities of dust and gas:
\begin{equation}
    Z = \frac{\Sigma_{\mathrm{d}}}{\Sigma_{\mathrm{g}}}.
\end{equation}
 We then compare our measured $Z$ with the best-fit formula for the critical $Z$ proposed by \citet{Lim2025}:
\begin{equation}
    \log(Z_{\mathrm{crit}}) = 0.1(\log(\mathrm{St}))^2 + 0.07(\log(\mathrm{St})) -2.36.
    \label{eq::z_crit}
\end{equation}
Note that this fitting formula was originally derived for the range $0.01 < \mathrm{St} < 1$, which does not overlap with the dust sizes adopted in our simulations. It remains to be seen how $Z_{\mathrm{crit}}$ varies for $\mathrm{St} > 1$, and hence, we extrapolate the curve up to $\mathrm{St} < 100$ to cover the broader parameter space relevant to our analysis. Our results in this regime rely on the extrapolation that moderately decoupled particles follow the same behavioral trend as marginally coupled ones. Verification of this assumption is beyond the scope of this work, and our findings should be interpreted accordingly.

\section{Results}\label{chap:Result}
\subsection{Low-mass flybys}\label{schap:Lowflyby}
In this section, the labels St$\sim$15, St$\sim$30, St$\sim$60, and St$\sim$100 correspond to the simulation IDs st15co, st30co, st60co, and st100co, respectively (see Table~\ref{table:setuptable}). 
\subsubsection{Disc Morphology}\label{sschap:DiscMorphology}

\begin{figure}
    \centering
    \includegraphics[width=\columnwidth]{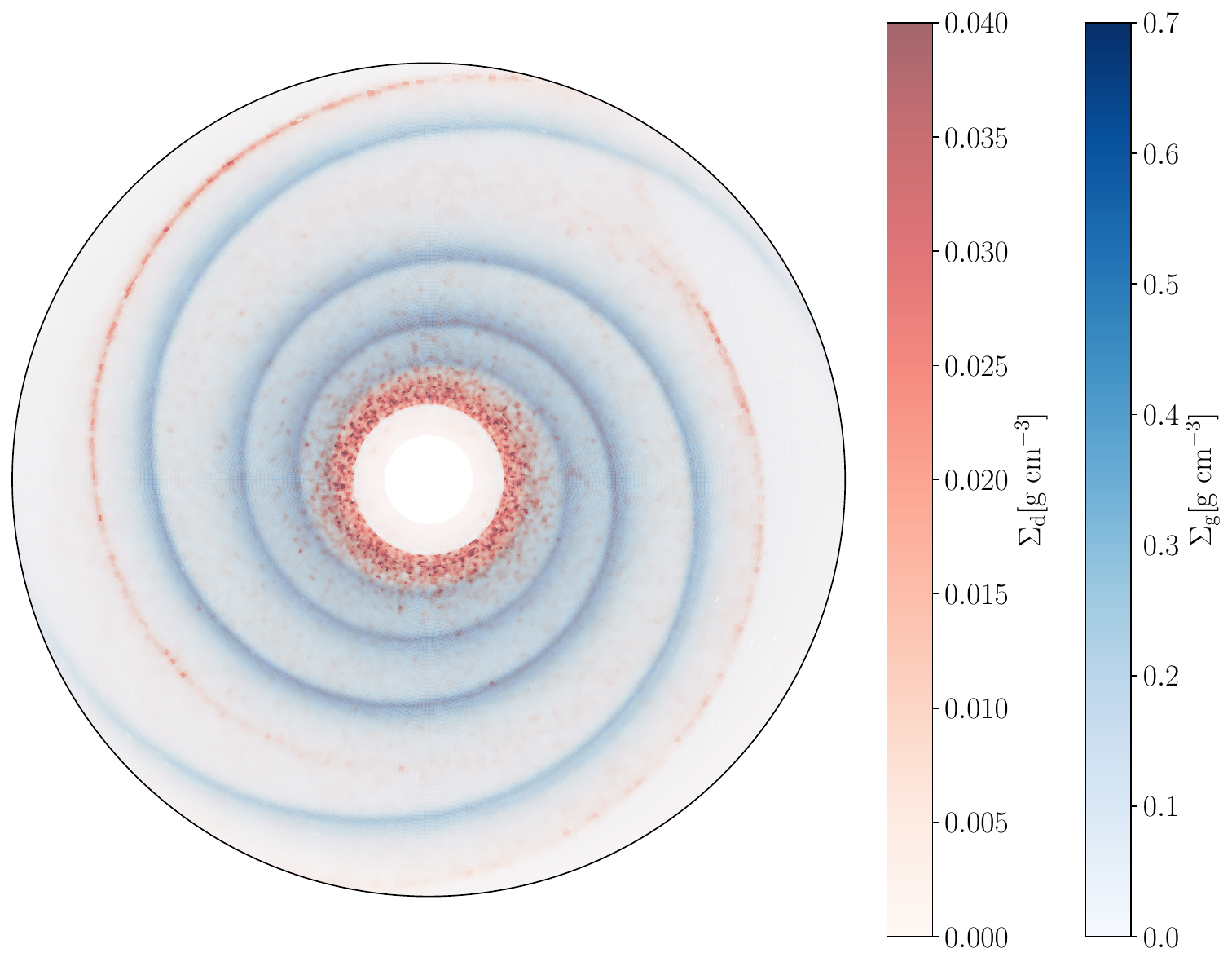}
    \caption{The polar projection of column density distributions for gas ($\Sigma_{\rm g}$, shown in blue) and dust ($\Sigma_{\rm d}$, shown in red) in the st15co simulation at $t = 8743$ yr highlights the differences between the two components. This visualization reveals an offset between the dust and gas spirals, indicating that the dust spirals do not coincide with the gas spirals.}
    \label{fig:spiral_offset} 
\end{figure}
\begin{figure}
    \centering
    \includegraphics[width=1\columnwidth]{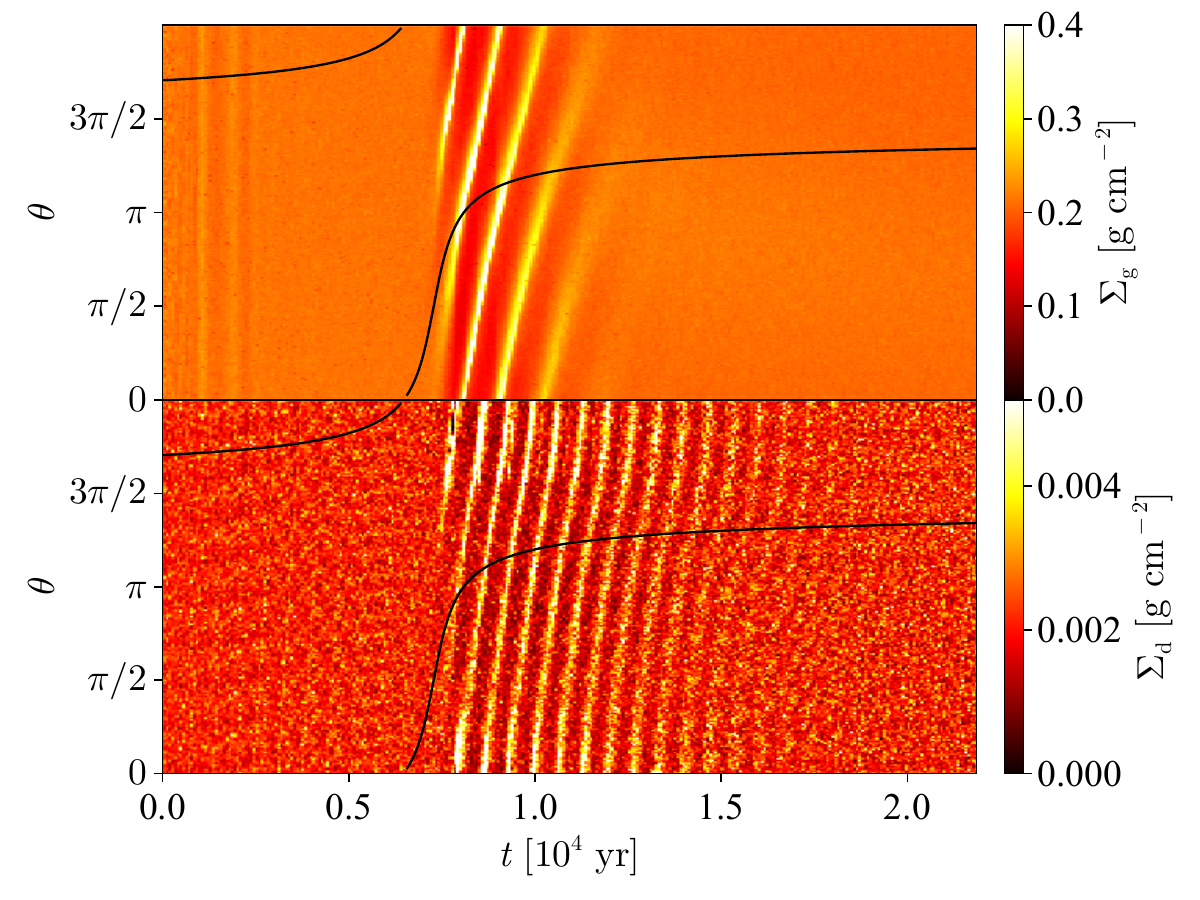}
    \caption{An azimuthal cut of low-mass coplanar case with initial $\rm St\sim 15$. We fix the radius at 75 au and calculate the column density of both gas and dust for every angle at each time step. The upper panel represents the azimuthal cut of the gas disc, while the lower panel represents the dust disc. The black line indicates the azimuthal position of the flyby. }
    \label{fig:Azimuthal_cut}
\end{figure}

We first present the evolution of the disc from a face-on perspective. The simulation images, extracted using \textsc{splash} \citep{Price2007}, are shown in \fref{fig:Low-mass_Splash_movie} for simulation IDs st15co, st30co, st60co, and st100co, as listed in \tref{table:setuptable}. Since the evolution of gaseous discs is similar across different Stokes numbers, we display only the gaseous disc for st15co to highlight the dynamical differences between the gaseous and dusty discs.  Our simulations show that while dust with $\rm St\approx15$ forms a well-defined inner ring due to efficient radial drift, the ring boundary becomes progressively more diffuse at higher Stokes numbers. This blurring occurs because larger grains (e.g., $\rm St \approx100$) are less coupled to the gas, leading to less efficient radial drift and trapping near the inner disc edge. This behaviour can be explained by the fact that dust particles with Stokes numbers close to unity experience faster radial drift and tend to pile up \citep{Gonzalez2017, Drazkowska2016}. Additionally, a spiral structure emerges following the flyby encounter. In the following discussion, arm 1 and arm 2 of spiral structures are labeled according to the order in which they are generated. A more detailed discussion of the spiral properties is provided in Sections \ref{sschap:PatternSpeed} and \ref{sschap:PitchAngle}. 

\fref{fig:AzimuthallyAveraged} shows the evolution of column density and eccentricity using azimuthally averaged radius–time profiles. In the gaseous disc, an increase in eccentricity is observed in the outer disc following the flyby perturbation, although this enhancement subsequently dissipates. In contrast, the outer region of the dusty disc exhibits a persistent rise in eccentricity, with higher Stokes numbers leading to an even longer duration of elevated eccentricity due to the reduced drag force. Additionally, a secondary increase in eccentricity is observed in the inner disc. In gaseous component, this inner-disc eccentricity increase is more pronounced than in dusty component; consequently, dust discs with lower Stokes numbers exhibit a greater increase in inner-disc eccentricity compared to those with higher Stokes numbers. When compared with \fref{fig:AzimuthallyAveragedSDP} and \fref{fig:AzimuthallyAveragedECC}, the outer regions of the discs display lower surface densities coupled with higher eccentricity growth, suggesting that the observed high eccentricity growth may be influenced by insufficient particle resolution. 

Figure~\ref{fig:WeightAveragedECC} details how the disc eccentricity evolves in time. The first panel shows the density-weighted eccentricity of discs undergoing a low-mass coplanar flyby, demonstrating that the gas exhibits higher eccentricity than the dust. The secondary increase, as previously mentioned, occurs following a small dip shortly after the perihelion passage. This dip coincides with the gas disc reaching its peak eccentricity, after which the gas eccentricity declines toward its pre-flyby level, while the dust eccentricity continues to rise and eventually surpasses that of the gas.

Figure \ref{fig:spiral_offset} presents the polar projection of column density distributions for gas (\(\Sigma_{\rm g}\), shown in blue) and dust (\(\Sigma_{\rm d}\), shown in red) within the context of the st15co simulation at a time \(t = 8743\) years.  Both gas and dust exhibit spiral structures, which are excited by the flyby encounter. The gas spirals are more pronounced and extend further into the inner disc compared to the dust spirals, a behaviour that is not related to the numerical resolution of the dust component (see Appendix~\ref{app:DustSpiralResolutionStudy}). There is a noticeable offset between the dust and gas spirals due to the dust being moderately coupled to the gas. This offset shows that the dust spirals do not coincide with the gas spirals.

\begin{figure}
    \centering
    \includegraphics[width=1\columnwidth]{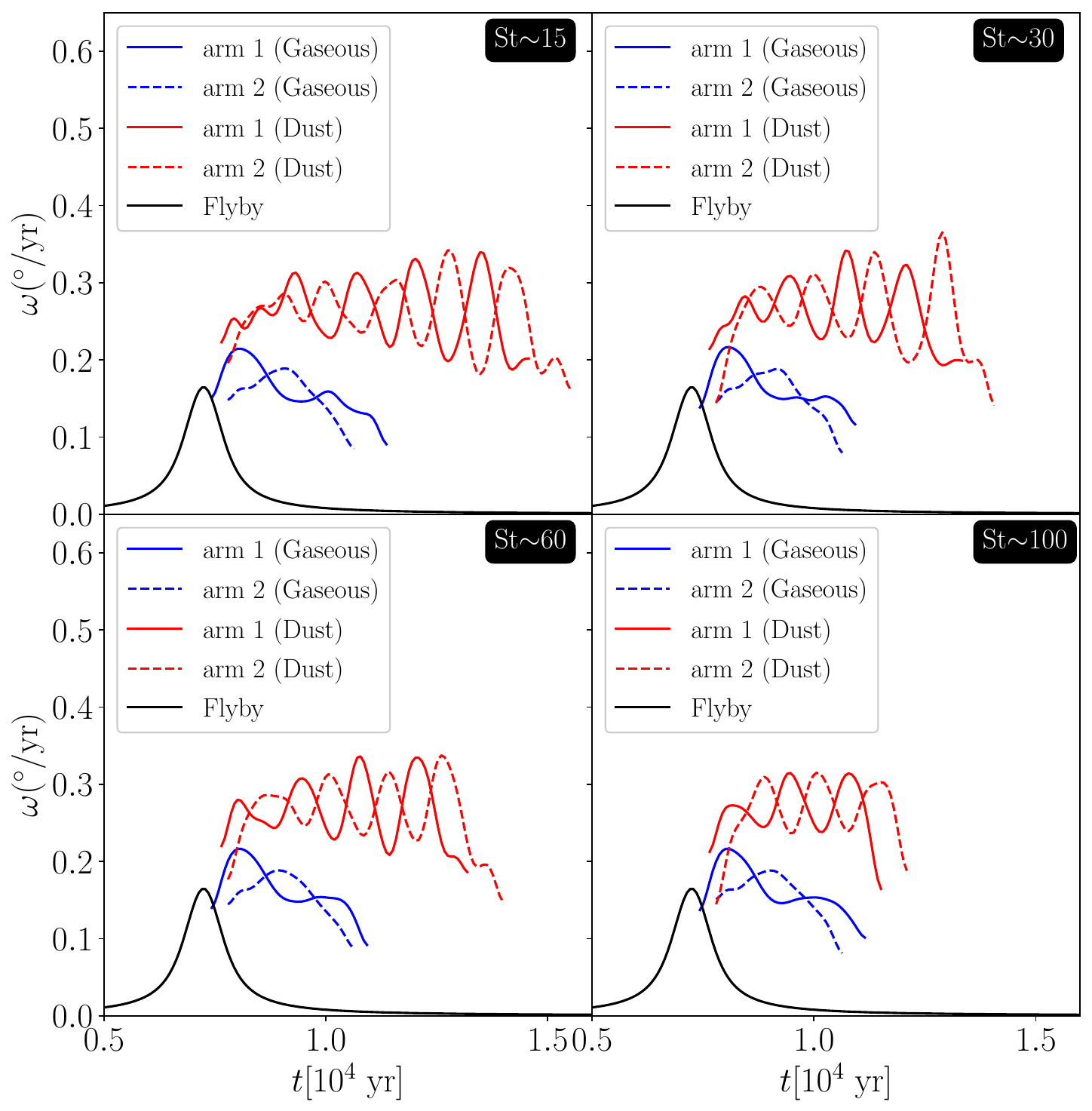}
    \caption{Evolution of the pattern speed ($\omega$) for two spiral arms in the gas and dust discs during a low-mass flyby encounter as a function of time ($t$) at $r_0 = 75$ au. Each panel shows the pattern speed evolution for four initial Stokes numbers (St), corresponding to the two gas and two dust spirals illustrated in \fref{fig:spiral_offset}. The solid black line indicates the angular velocity of the flyby perturber. Arms 1 and 2 are labeled chronologically according to their formation time.}
    \label{fig:PatternEvolution_coplanar}
\end{figure}

\begin{figure}
    \centering
    \includegraphics[width=1\columnwidth]{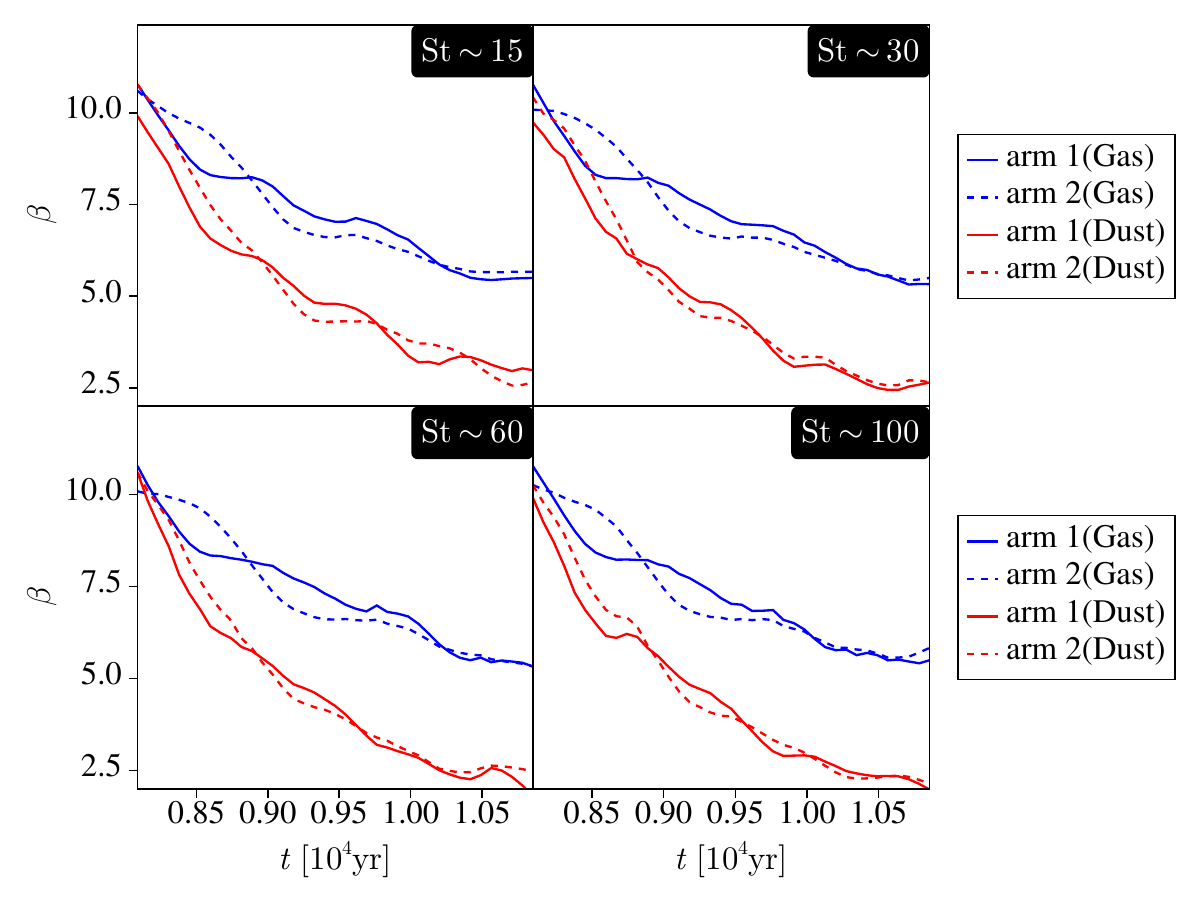}
    \caption{Evolution of the pitch angle of spirals in both gaseous and dusty discs during a low-mass coplanar flyby encounter. The blue lines represent the evolution of pitch angle $\beta$ for gaseous spirals, while the red lines represent this evolution for dust spirals. The definitions of arm 1 and arm 2 are identical to those in Figure \ref{fig:PatternEvolution_coplanar}. Each panel represents four initial Stokes numbers (St).}
    \label{fig:pitch_angle_coplanar}
\end{figure}

\begin{figure}
    \centering
    \includegraphics[width=1\columnwidth]{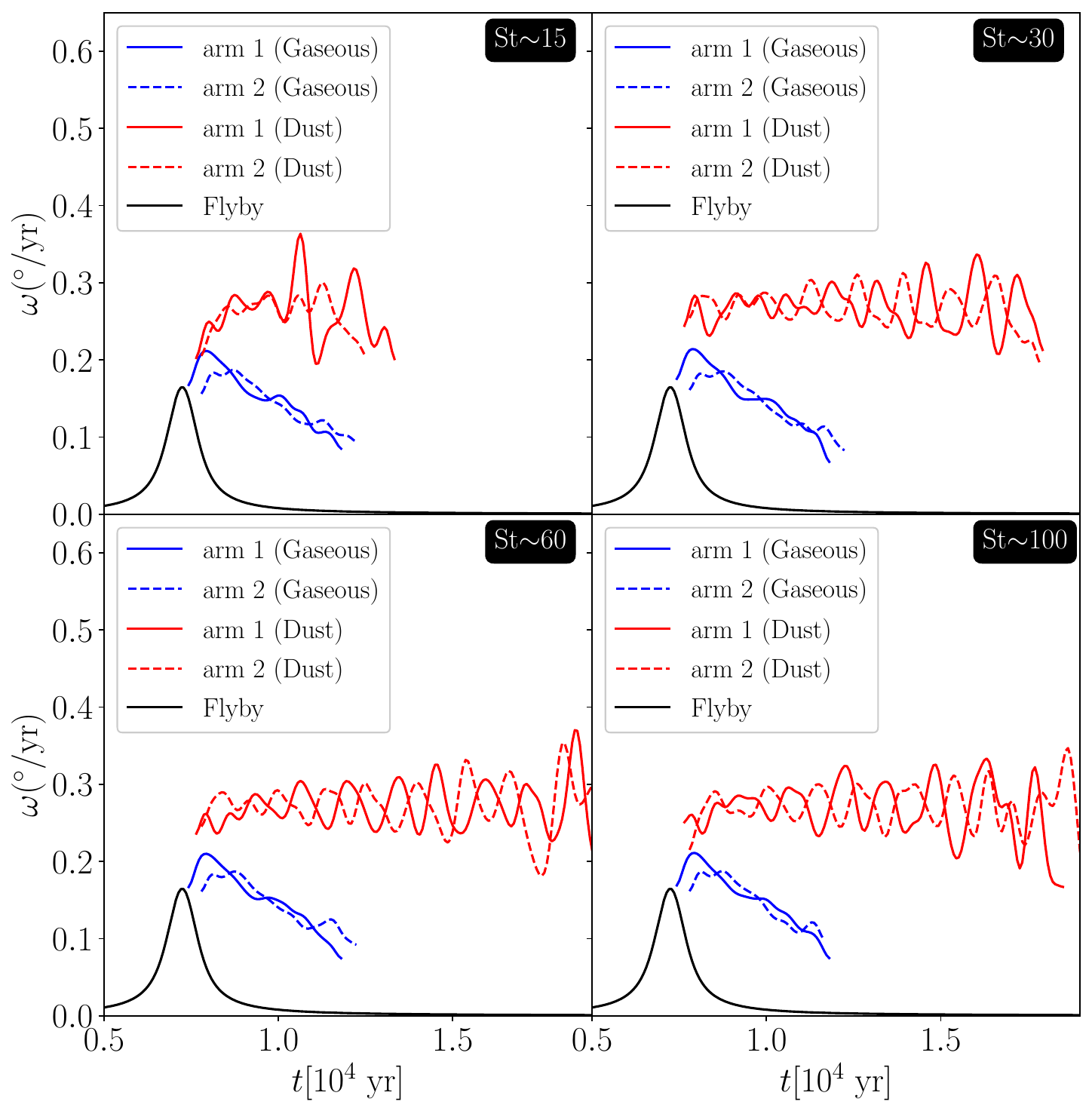}
    \caption{Same as Figure~\ref{fig:PatternEvolution_coplanar} but for an inclined flyby encounter.}
    \label{fig:PatternEvolution_inclined}
\end{figure}

\begin{figure}
    \centering
    \includegraphics[width=1\columnwidth]{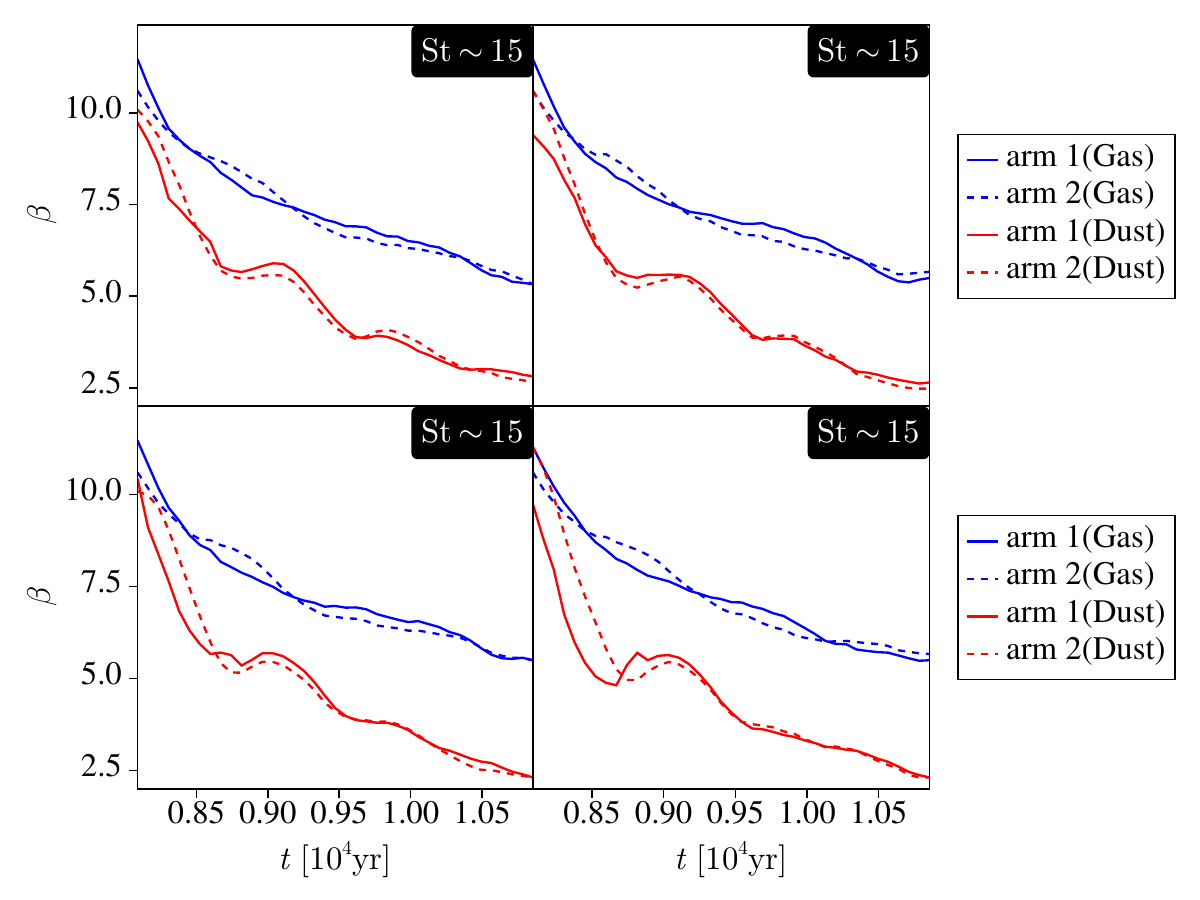}
    \caption{Same as Figure~\ref{fig:pitch_angle_coplanar} but for an inclined flyby encounter.}
    \label{fig:pitch_angle_inclined}
\end{figure}

\begin{figure*}
    \centering
    \includegraphics[width=2\columnwidth]{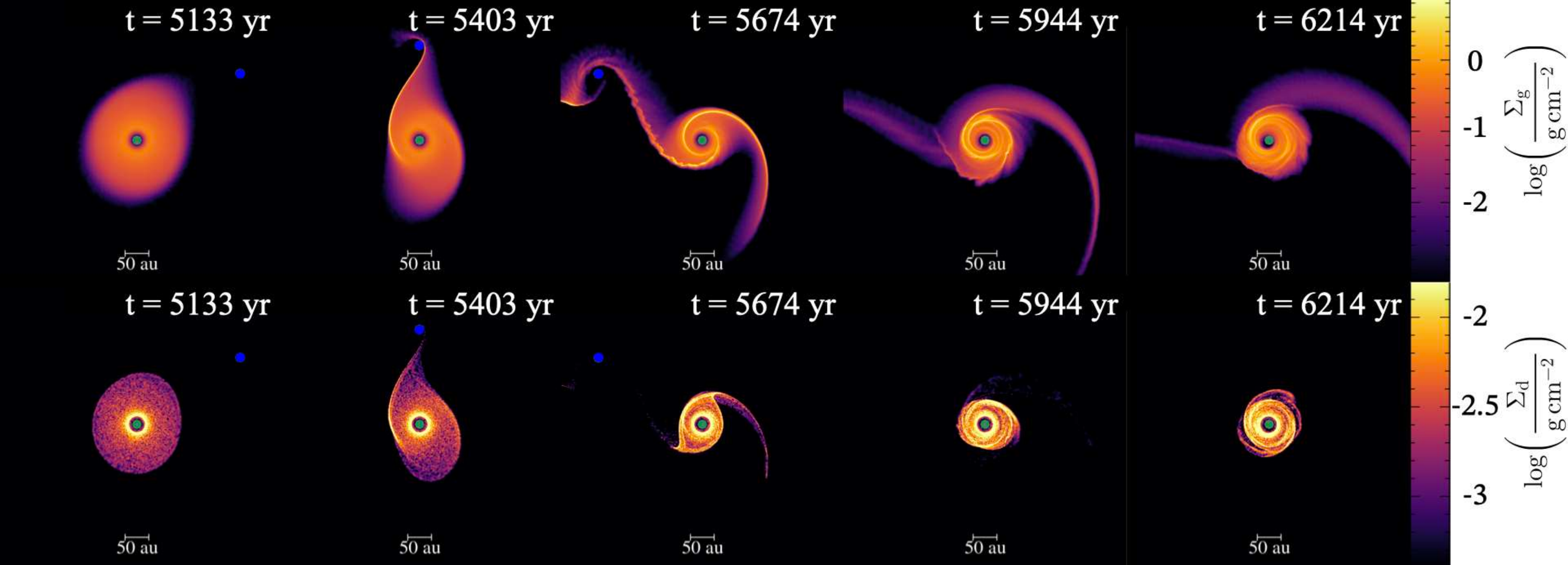}
    \caption{The evolution of density in gaseous and dusty discs in st15co-eqmass. The equal-mass flyby shows a more destructive influence on the disc. After the flyby reaches the perihelion, the radius of the disc shrinks to $\sim 50$ au.}
    \label{fig:Equal-mass_Splash_movie}
\end{figure*}

\subsubsection{Pattern speed}\label{sschap:PatternSpeed}
Figure \ref{fig:Azimuthal_cut} shows an azimuthal cut of the disc surface density for st15co used to calculate the pattern speed of the spiral arms. The solid black line represents the azimuthal velocity of the flyby from the primary star's point of view as time evolves, with the azimuthal angle of the periastron point being $90^\circ$. The dusty spiral has a longer lifetime than the gaseous spiral before dissipating. It should be noted that estimating the exact lifetime of spirals in SPH simulations can be challenging, as the result strongly depends on the resolution of the simulation \citep{Smallwood2023}.

Figure \ref{fig:PatternEvolution_coplanar} shows the temporal evolution of the spiral pattern speed, which we estimate from azimuthal cuts of the disc. Initially, the pattern speed is comparable to the azimuthal velocity of the flyby at perihelion. Subsequently, the gaseous and dusty components diverge: the pattern speed of the gaseous spirals steadily decreases, while the speed of the dusty spirals oscillates quasi-periodically around an average value of $\sim 0.3^{\circ}/\mathrm{yr}$. These oscillations form wave-like patterns with a clear phase offset between the two arms. Despite this variability, the time-averaged pattern speed of each dusty spiral remains nearly constant throughout the simulation.

\subsubsection{Pitch angle}\label{sschap:PitchAngle}

Figure~\ref{fig:pitch_angle_coplanar} illustrates the evolution of the pitch angle ($\beta$) for spirals in both gaseous and dust disc during a \textit{low-mass coplanar flyby}. The four subplots correspond to different initial Stokes numbers ($\text{St} \sim  15, 30, 60, 100$), representing varying levels of dust-gas coupling. The gas spiral arms are shown in blue, while the dust spiral arms are depicted in red, with solid and dashed lines distinguishing arm 1 and arm 2.

The pitch angles of both gas and dust spirals start similarly but gradually decrease over time, indicating that both spiral types become more tightly wound as the system evolves. Notably, the pitch angle of the dust spirals decreases faster than that of the gas spirals, indicating that dust spirals wind up faster. However, as discussed previously, gas spirals tend to dissipate before they become tightly wound.

\subsection{Effects of flyby inclination}\label{sschap:InclinedFlyby}
In this section, we present the results of our simulations modelling a flyby encounter featuring an orbital inclination of $45^\circ$. The labels St$\sim$15, St$\sim$30, St$\sim$60, and St$\sim$100 correspond to the simulation IDs st15in, st30in, st60in, and st100in, respectively (see Table~\ref{table:setuptable}). We note that the evolution of the inclined flyby simulation from a face-on perspective does not differ significantly from the coplanar cases as shown in \fref{fig:Low-mass_Splash_movie}. The overall pattern of azimuthal-averaged eccentricity evolution is similar to \fref{fig:AzimuthallyAveragedECC}, but with marginally reduced values. However, the evolution of the density-weighted eccentricity for the dust components, shown in Figure~\ref{fig:WeightAveragedECC}, deviates from that of the coplanar case: rather than increasing after the initial minor dip, it remains nearly constant throughout the evolution. 

\subsubsection{Pattern speed and pitch angle}
Figure \ref{fig:PatternEvolution_inclined} illustrates the temporal evolution of the spiral pattern speed. The overall evolution and magnitude of the pattern speed in the inclined flyby scenario closely match those observed in the coplanar flyby case. This consistency suggests that the spiral pattern speed is largely independent of both the Stokes number of the disc particles and the inclination angle of the flyby perturber. However, when examining the fluctuations, the trend is clear for the gas but less so for the dust. For the gas, the oscillations in pattern speed are less pronounced in the inclined flyby than in the coplanar case. For the dust, we do not find a comparably clear reduction in the fluctuation amplitude over time in Fig.~\ref{fig:PatternEvolution_inclined}, and any difference between the two configurations appears weaker.

Figure~\ref{fig:pitch_angle_inclined} depicts the temporal evolution of the spiral pitch angle. The evolution of the pitch angle in discs perturbed by an inclined flyby follows a trend closely aligned with that seen in coplanar flyby cases. A similar behaviour is found for the gas pitch angle: the inclined case shows diminished temporal variations compared to the coplanar encounter. In contrast, for the dust the reduction in the oscillatory behaviour is not clearly evident in Fig.~\ref{fig:pitch_angle_inclined}, and the time-dependent variations are broadly comparable between the two geometries.

\subsection{Equal mass flyby}\label{schap:EqualFlyby}
\begin{figure*}
    \begin{subfigure}[b]{0.45\textwidth}
    \centering
    \includegraphics[width=1\columnwidth]{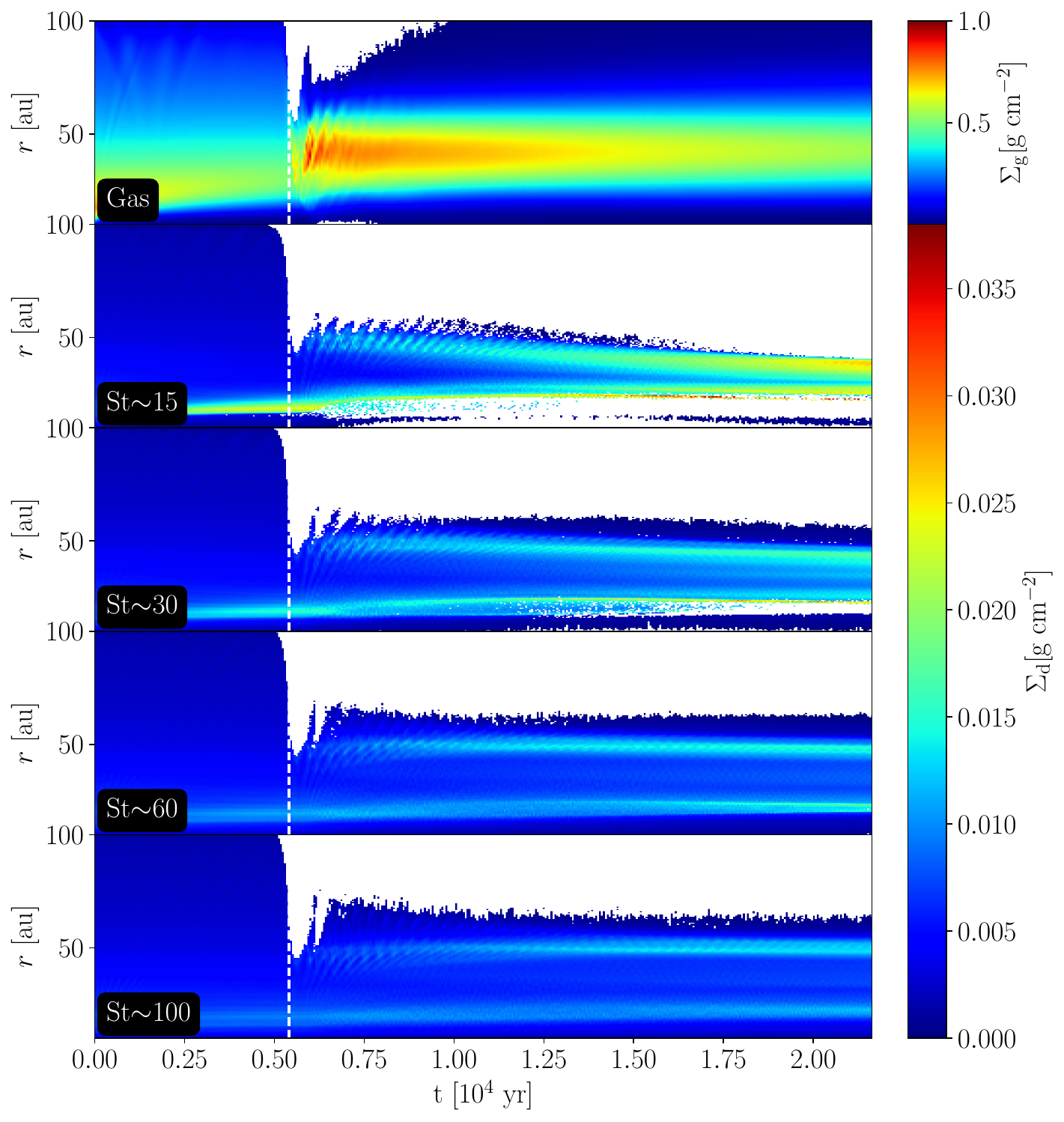}
    \caption{Azimuthally-averaged column density profile}
    \label{fig:AzimuthallyAveragedSDPMassive}
    \end{subfigure}
    \begin{subfigure}[b]{0.45\textwidth}
    \centering
    \includegraphics[width=1\columnwidth]{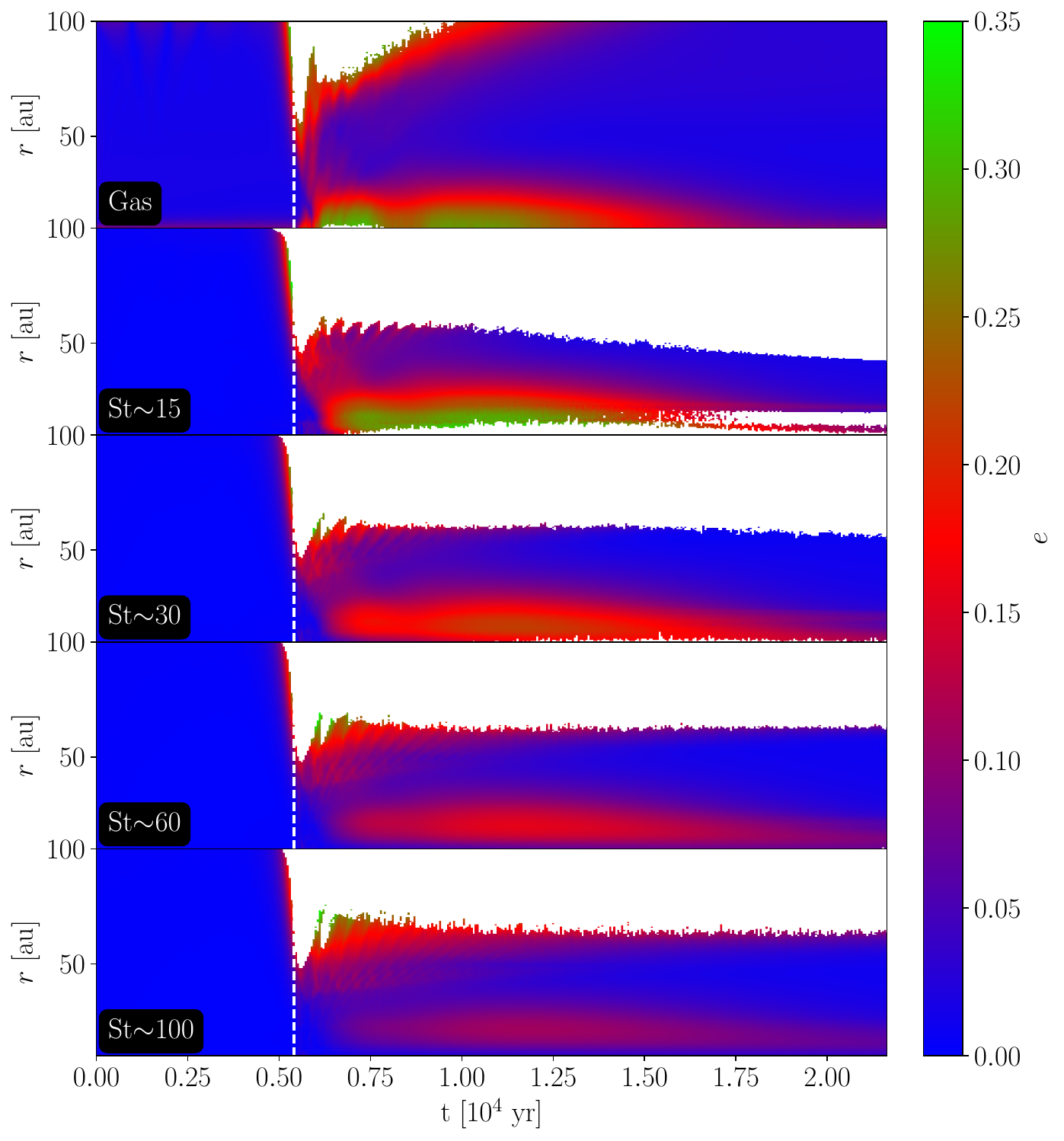}
    \caption{Azimuthally-averaged eccentricity profile}
    \label{fig:AzimuthallyAveragedECCMassive}
    \end{subfigure}
    \caption{Evolution of the azimuthally-averaged column density and eccentricity profiles in a radial-time grid with a coplanar equal-mass flyby. The white empty region corresponds to a density of zero at that radius. The white dashed lines represent the time when the flyby arrived at perihelion.}
    \label{fig:AzimuthallyAveragedMassive}
\end{figure*}
We performed a series of equal-mass flyby simulations, exploring initial averaged Stokes numbers ($\mathrm{St}$) of 15, 30, 60, and 100 for coplanar encounters. In addition, we included an inclined encounter with $\mathrm{St} \approx 60$. In this section, the labels St$\sim$15, St$\sim$30, St$\sim$60, and St$\sim$100 correspond to the simulation IDs st15co-eqmass, st30co-eqmass, st60co-eqmass, and st100co-eqmass, respectively (see Table~\ref{table:setuptable}). 

\subsubsection{Disc morphology}\label{sschap:DiscMorphologyEq}

\fref{fig:Equal-mass_Splash_movie} illustrates the evolution of the disc induced by a coplanar equal-mass flyby in the simulation labelled st15co-eqmass. Following the flyby encounter, a bridge connecting the primary and secondary sink particles emerges, consistent with the findings of \cite{Cuello2019}. Moreover, a pair of spiral arms forms shortly after the flyby reaches perihelion, a feature that proved challenging to trace in previous attempts \citep{Smallwood2023}. The first-formed spiral arm extends toward the flyby. Consistent with previous results, the more massive flyby companion more heavily truncates the gas and dust.

As shown in \fref{fig:AzimuthallyAveragedSDPMassive}, the disc is strongly perturbed by the massive flyby, resulting in mass loss and a reduction in radius. The truncation of radius will be discussed in Section \ref{sschap:DiscTruncation}. Dust particles are concentrated near the outer edge of the truncated disc, forming a tightly wound spiral structure with a ring-like appearance, and gradually migrate inward. These ring-like features in dust discs with lower Stokes numbers migrate faster than those associated with higher Stokes numbers.

Moreover, \fref{fig:AzimuthallyAveragedECCMassive} shows the eccentricity growth in the inner disc. Both gas and dust components exhibit larger eccentricities compared to scenarios with less massive flyby companions. To further investigate the eccentricity evolution, the density-weighted eccentricity is extracted and presented in the bottom panel of \fref{fig:WeightAveragedECC}. The evolution of gas eccentricity in discs perturbed by an equal-mass flyby shows a much stronger increase compared to those perturbed by a low-mass flyby. In the dust disc, the eccentricity increases immediately after the flyby reaches perihelion, then drops as a dip, before returning to the pre-dip eccentricity value without further increase, resembling the behaviour observed in inclined cases. The post-dip decrease in eccentricity indicates that the recircularization of the disc in these simulations is more pronounced than in the low-mass simulations. Furthermore, the eccentricity evolution in the disc perturbed by an inclined equal-mass flyby demonstrates a more subdued recircularization process.

\subsubsection{Disc truncation}\label{sschap:DiscTruncation}
\begin{figure}
    \centering
    \includegraphics[width=\columnwidth]{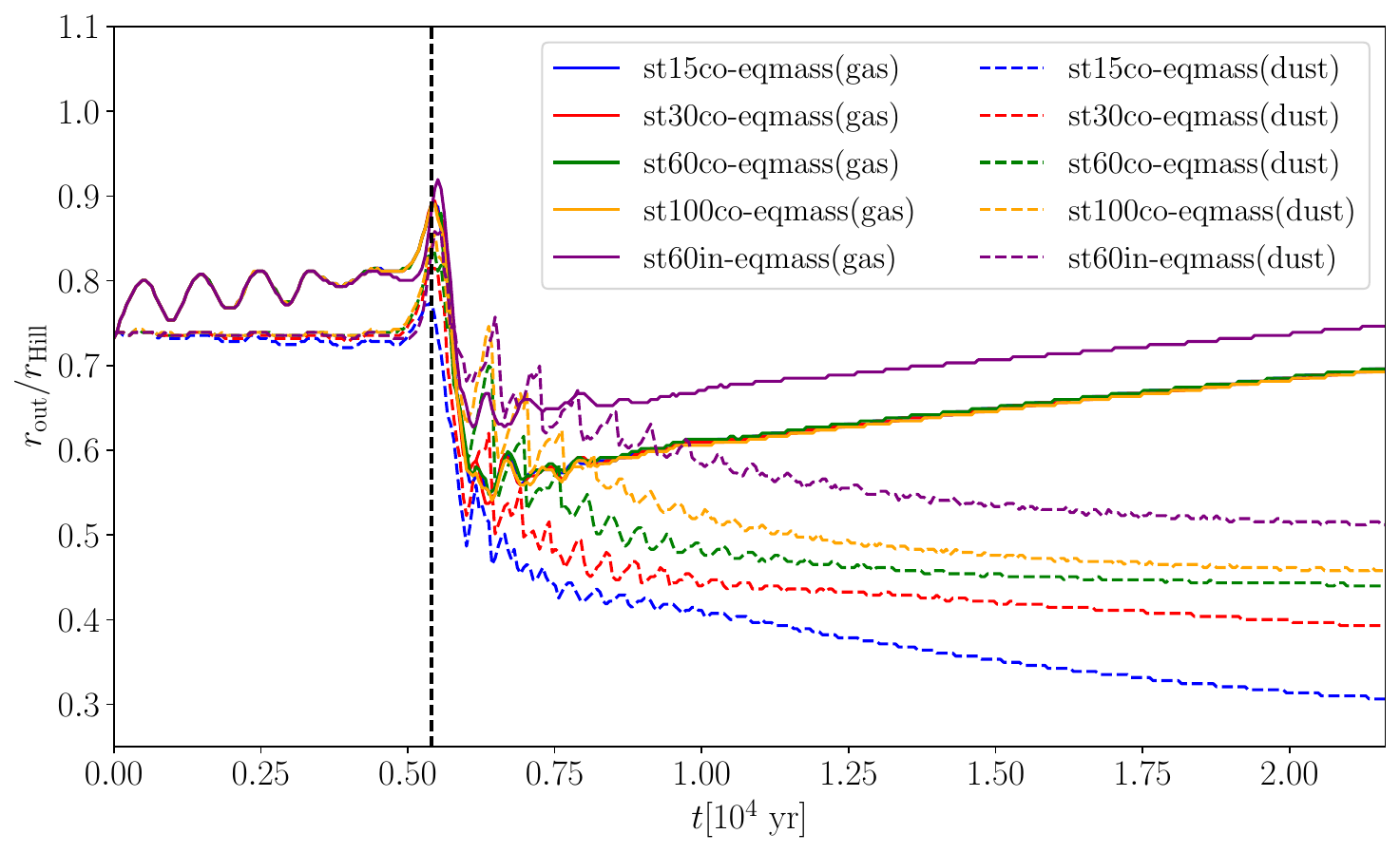}
    \caption{The evolution of the disc radius during massive flyby encounters. The vertical black dashed line indicates the time when the massive flyby reaches perihelion. The solid line represents the radius of the gas discs, while the non-vertical dashed line represents the radius of a dust disc. The Hill radius has normalized the radius.}
    \label{fig:DiscTruncation}
\end{figure}

We estimate the truncation radius $r_{\mathrm{out}}$ after an equal-mass flyby encounter by identifying the radius at which the azimuthally averaged column density falls below the mean column density, as shown in \fref{fig:DiscTruncation}. The truncation radius has been normalized by Hill radius:
\begin{equation}
    r_{\mathrm{Hill}} \approx r_{\mathrm{p}} \left(\frac{M_2}{3M_1}\right)^{1/3} \approx 138\,\mathrm{au}
\end{equation}
Prior to the flyby, gas discs undergo radial expansion, while the radius of dust discs remains unchanged. After the encounter, both gas and dust discs are truncated to approximately $0.55r_{\mathrm{Hill}}$ for a coplanar flyby and to $0.65r_{\mathrm{Hill}}$ for an inclined flyby. Dust discs, however, continue to  shrink, with the largest post-flyby radius observed in the coplanar equal-mass simulation st100co-eqmass. The difference of shrinking among different Stokes numbers of dust discs can be attributed to the weaker coupling between gas and dust at higher Stokes numbers, which reduces the drag force. Over time, gas discs continue to expand outward, whereas dust discs gradually decrease in size. The smaller the grain size, the faster the disc shrinks. By contrast, the evolution of discs influenced by an inclined massive flyby exhibits a similar but weaker effect. In contrast to the observed truncation radii, \cite{Bhandare2016} proposed a formula for the tidal truncation radius, averaged over inclination angles, which yields a value of approximately $0.52r_{\mathrm{Hill}}$, for our disc setup. This estimate is slightly smaller than our measured truncation radius, but the difference remains acceptable.

\begin{figure}
    \centering
    \includegraphics[width=1\columnwidth]{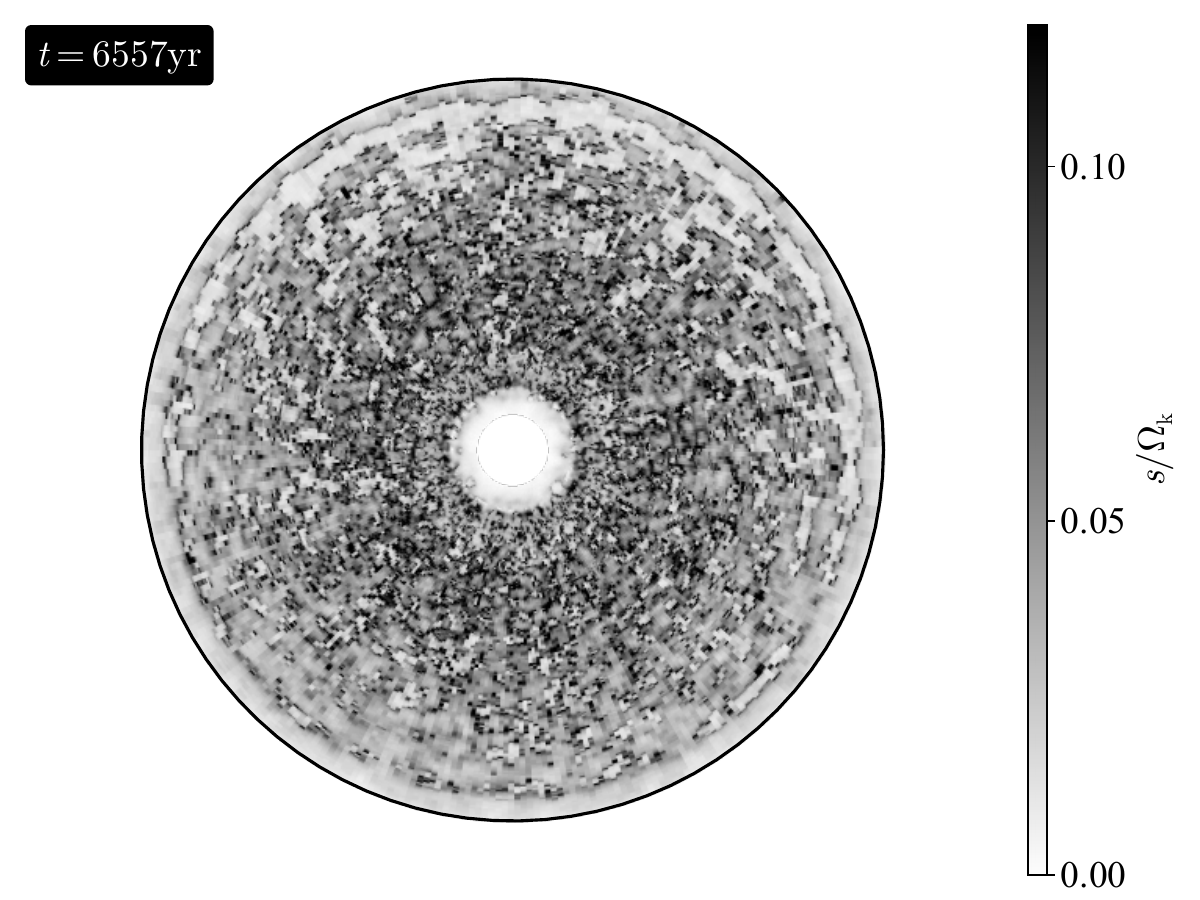}
    \includegraphics[width=1\columnwidth]{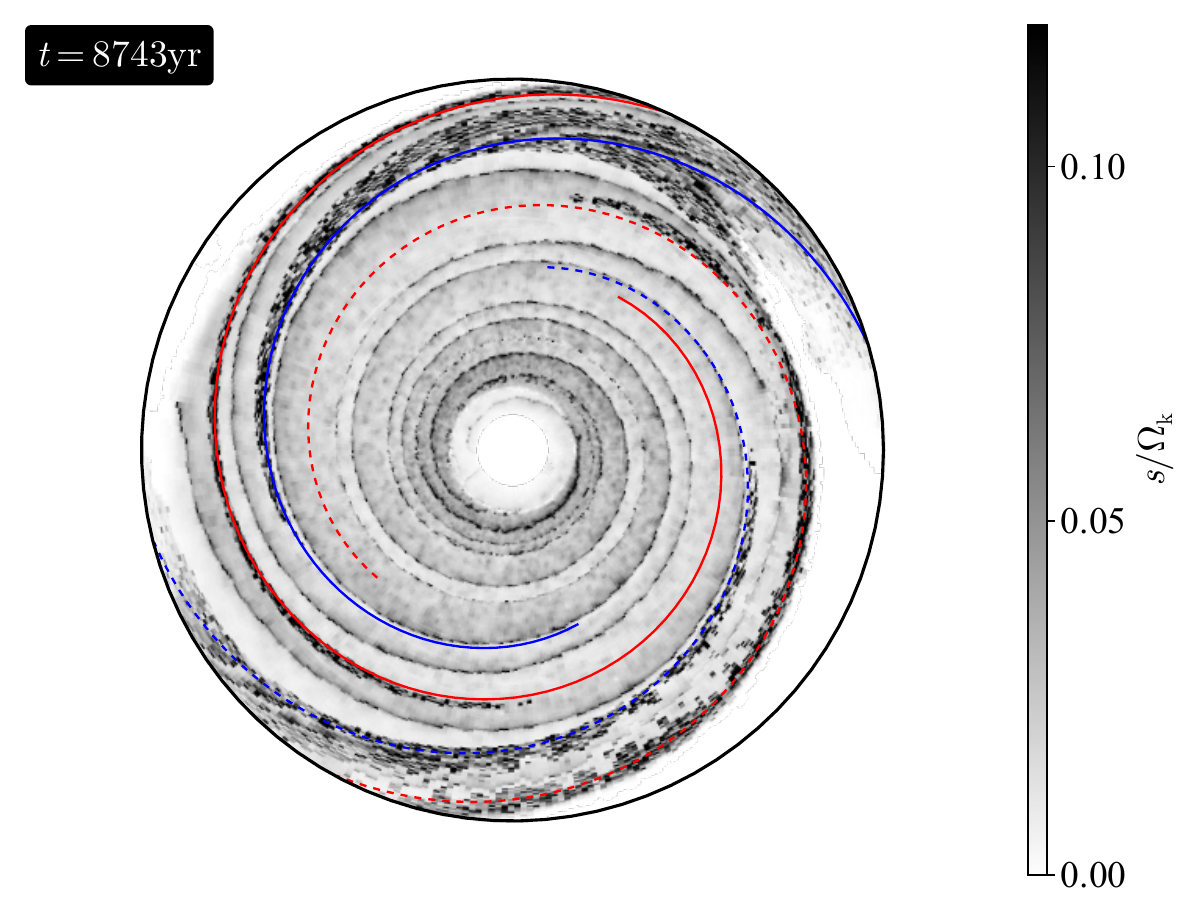}
    \caption{Spatial distribution of the streaming instability growth rate ($s/\Omega_{\rm k}$) at two snapshots: prior to the flyby's periastron passage (top, $t=6557\, \rm yr$) and after (bottom, $t=8743\, \rm yr$) in st15co. Overplotted on the bottom panel are the locations of the spiral arms. Solid lines denote arm 1 and dashed lines denote arm 2, with blue for gas and red for dust.}
    \label{fig:Faceon_growthrate}
\end{figure}

\begin{figure*}
    \begin{subfigure}[b]{0.45\textwidth}
    \centering
    \includegraphics[width=1\columnwidth]{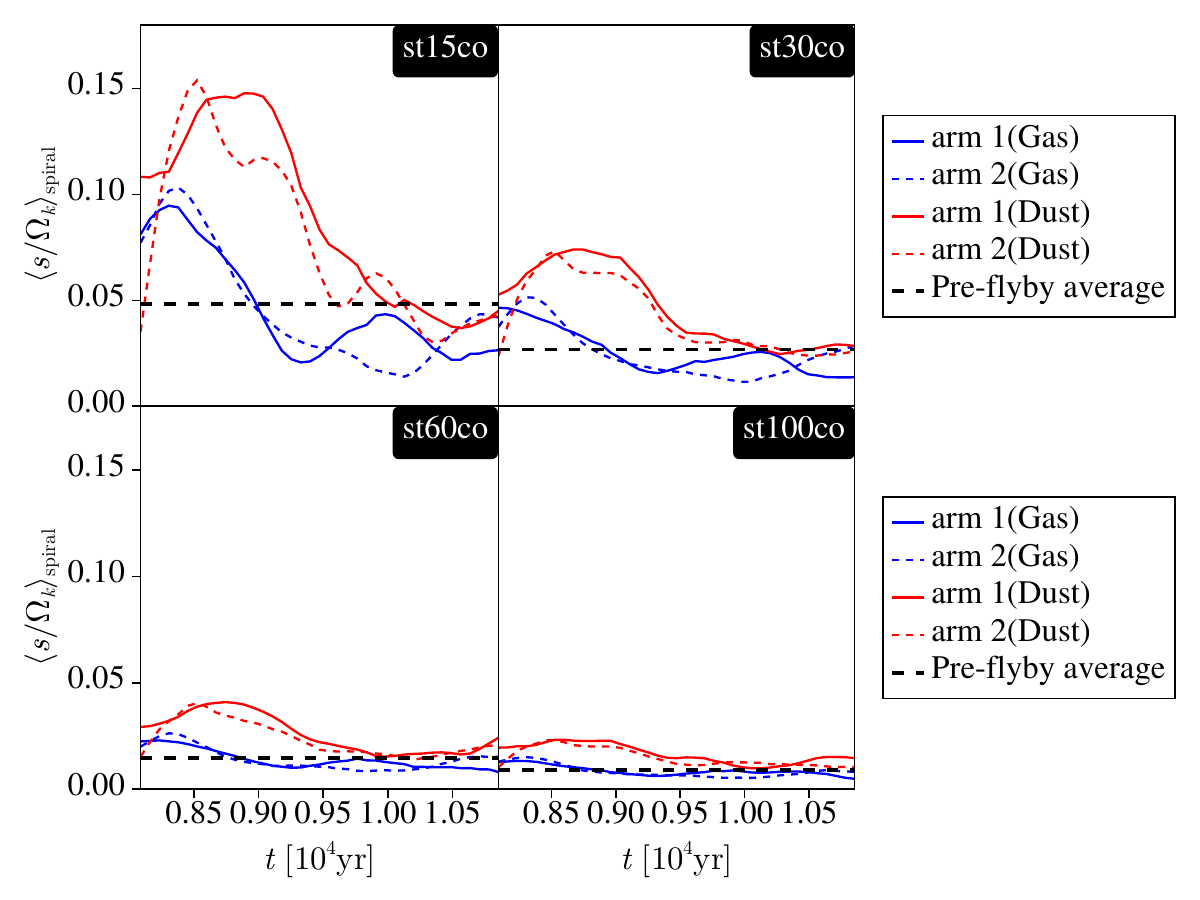}
    \caption{Low-mass coplanar}
    \label{fig:AveragedSpiralGRSIco}
    \end{subfigure}
    \begin{subfigure}[b]{0.45\textwidth}
    \centering
    \includegraphics[width=1\columnwidth]{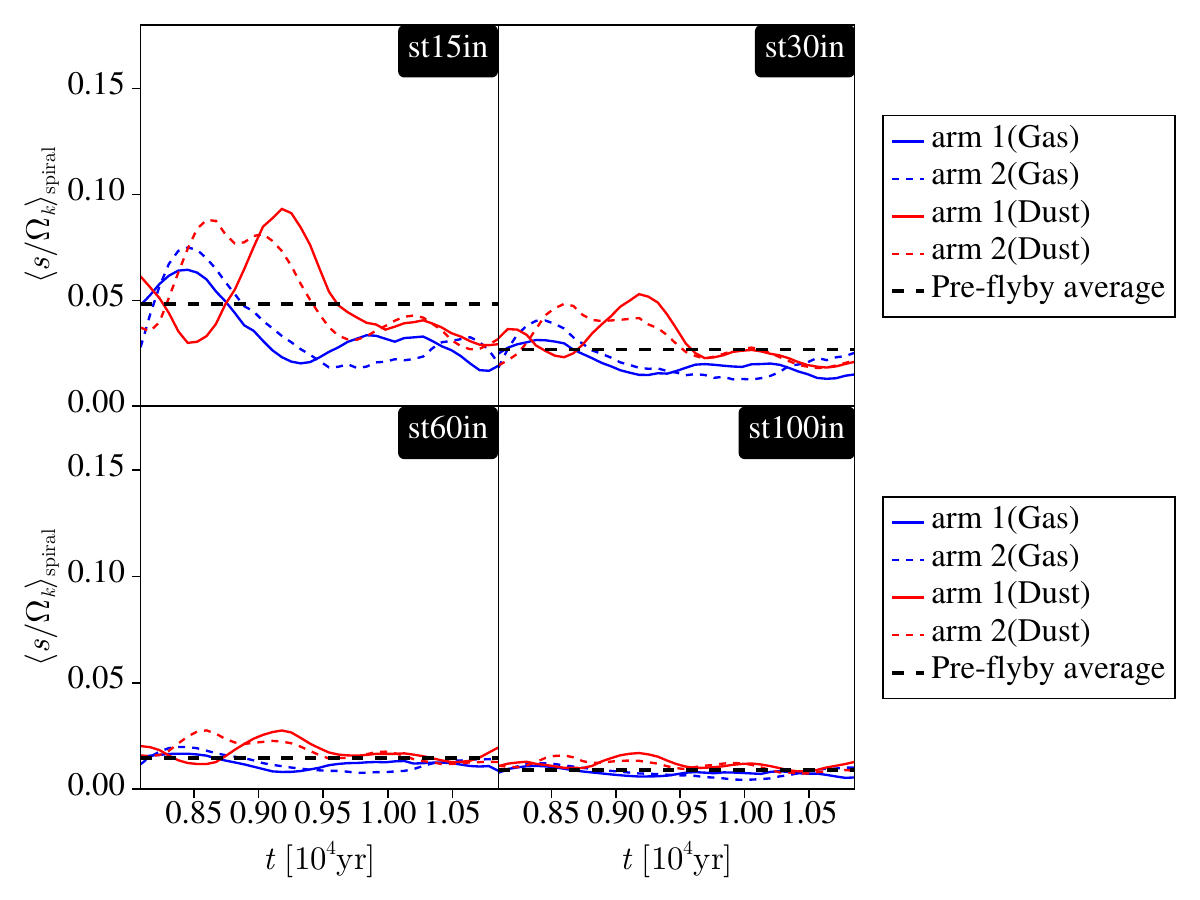}
    \caption{Low-mass inclined}
    \label{fig:AveragedSpiralGRSIin}
    \end{subfigure}
    \caption{Evolution of growth rate of streaming instability along the spirals in both gaseous and dust discs during a low-mass flyby encounter. The blue lines represent the evolution of growth rate $s/\Omega_{\mathrm{k}}$ for gaseous spirals, while the red lines represent those for dust spirals. The black dashed line represents the pre-flyby average growth rate of the disc, measured at $t\sim0.95 t_{\mathrm{p}}$. The definition of arm 1 and arm 2 is identical to the definition in \fref{fig:PatternEvolution_coplanar}. Each sub-panel represents four initial Stokes numbers (St).}
    \label{fig:AveragedSpiralGRSI}
\end{figure*}
\section{Application to Streaming Instability}\label{schap:ApplicationSI}

\begin{figure*}
    \begin{subfigure}[b]{0.3\textwidth}
    \centering
    \includegraphics[width=1\columnwidth]{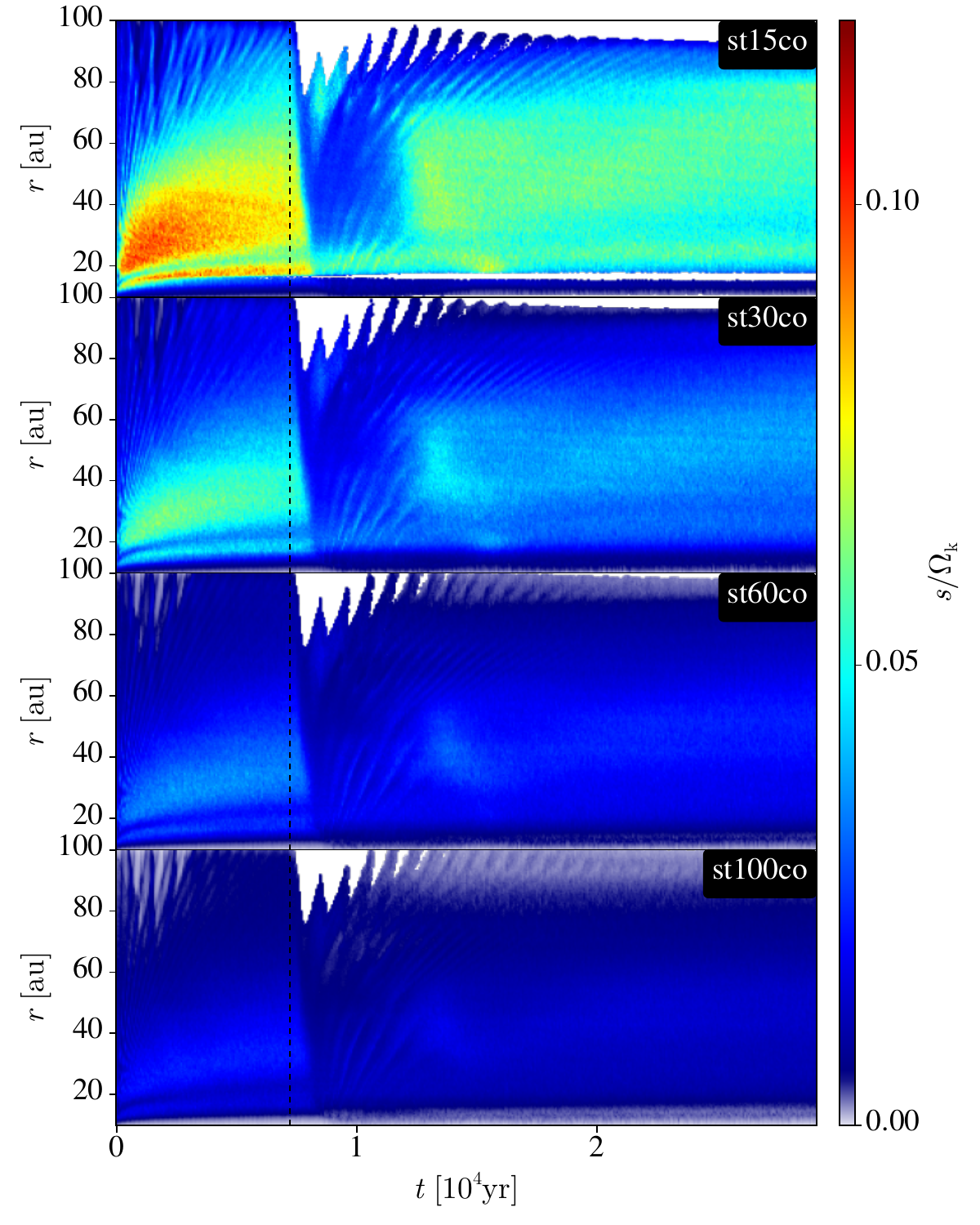}
    \caption{Low-mass coplanar flyby}
    \label{fig:AzimuthallyAveragedGrowthRatecolomass}
    \end{subfigure}
    \begin{subfigure}[b]{0.3\textwidth}
    \centering
    \includegraphics[width=1\columnwidth]{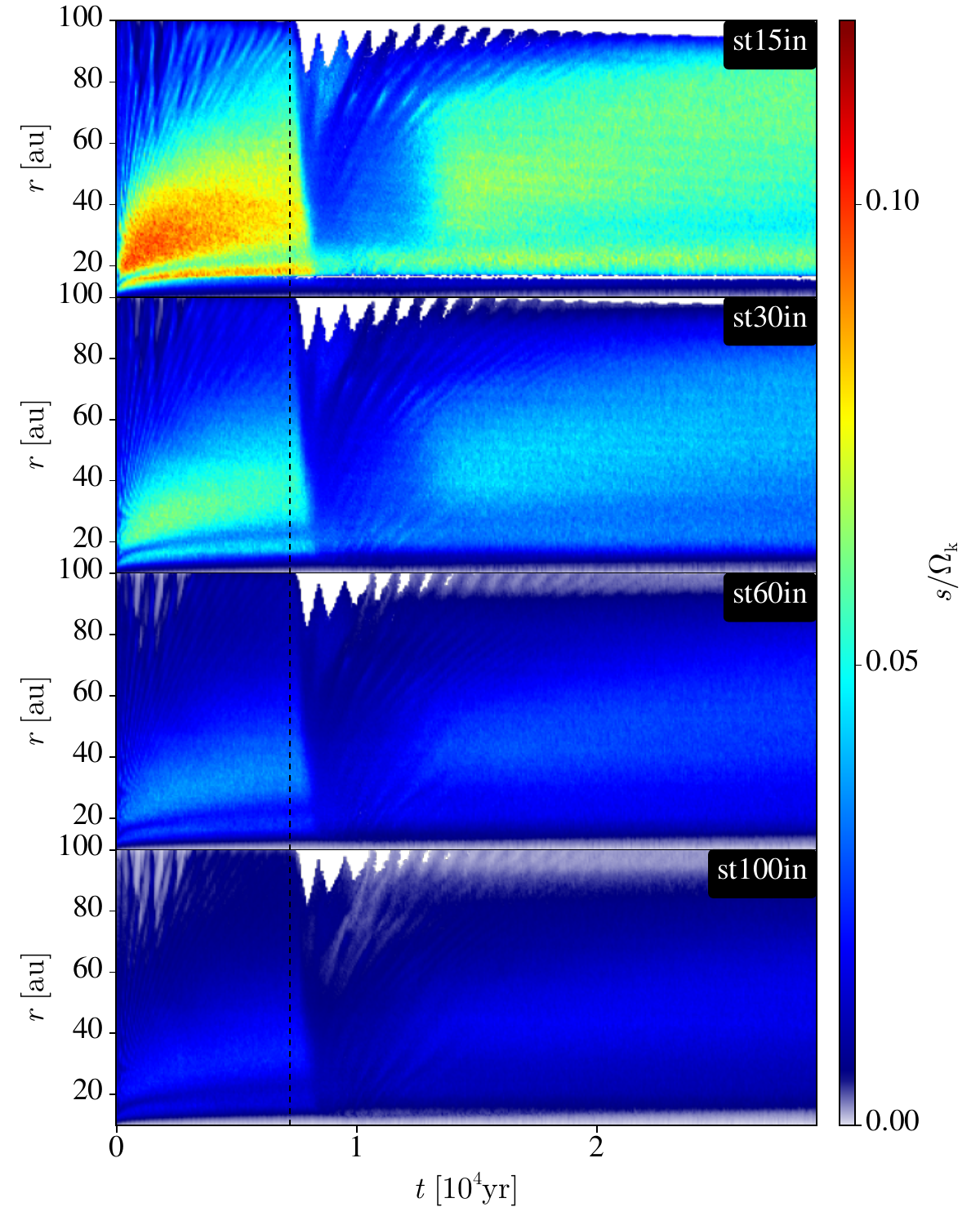}
    \caption{Low-mass inclined flyby}
    \label{fig:AzimuthallyAveragedGrowthRateinlomass}
    \end{subfigure}
    \begin{subfigure}[b]{0.3\textwidth}
    \centering
    \includegraphics[width=1\columnwidth]{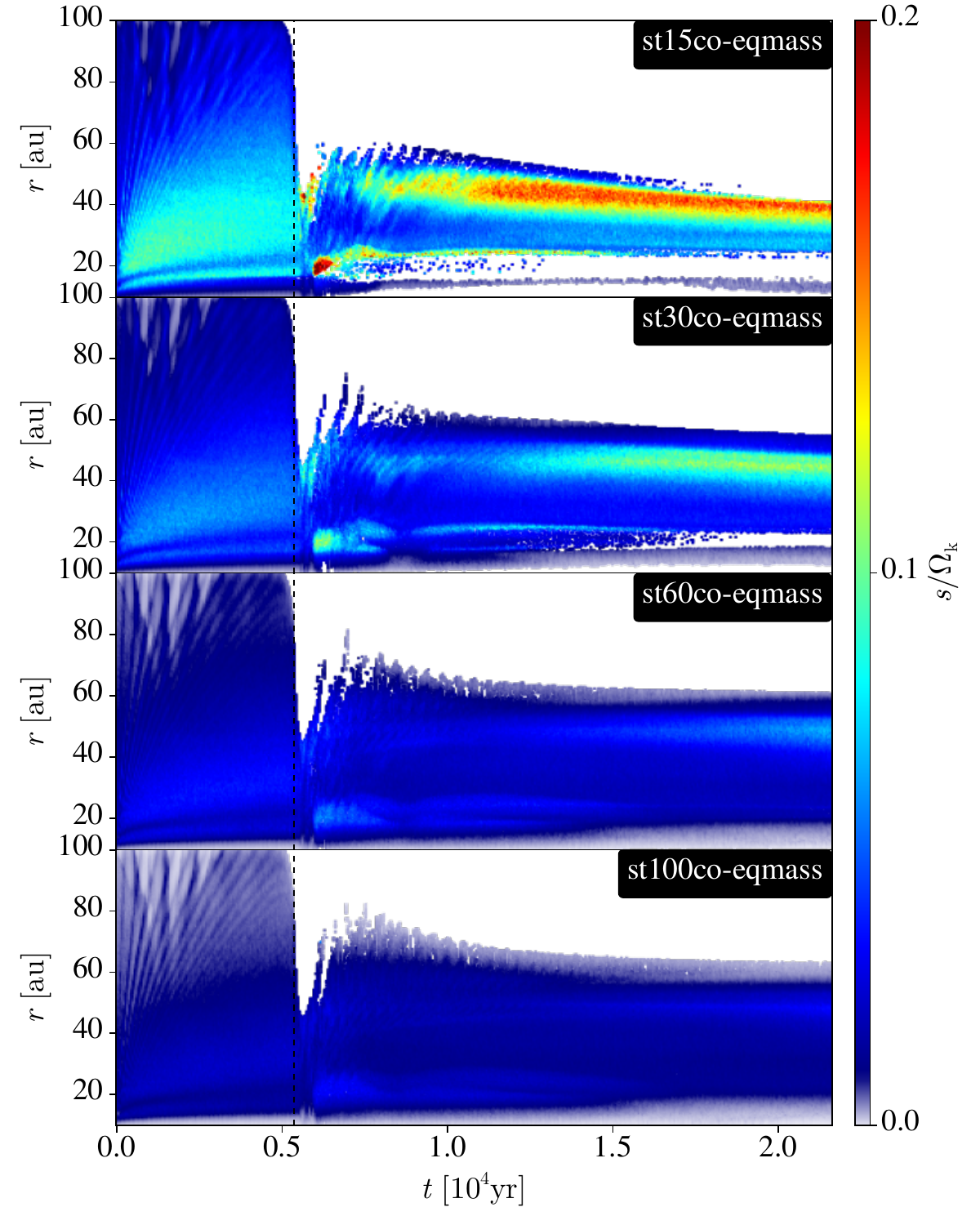}
    \caption{Equal-mass coplanar flyby}
    \label{fig:AzimuthallyAveragedGrowthRatecoeqmass}
    \end{subfigure}
    \caption{Evolution of the azimuthally-averaged growth rate ($s/\Omega_{\rm k}$) of streaming instability as a function of radius $r$ and time $t$ for three distinct flyby scenarios: (a) a low-mass coplanar flyby, (b) a low-mass inclined flyby, and (c) an equal-mass coplanar flyby. Each sub-panel denotes a different initial Stokes number of the dust. The vertical black dashed line marks the moment of periapsis passage.}
    \label{fig:AzimutahllyAveragedGrowthRate}
\end{figure*}

\begin{figure}
    \centering
    \hfill
    \begin{subfigure}[t]{0.97\columnwidth}
    \centering
    \includegraphics[width=\linewidth]{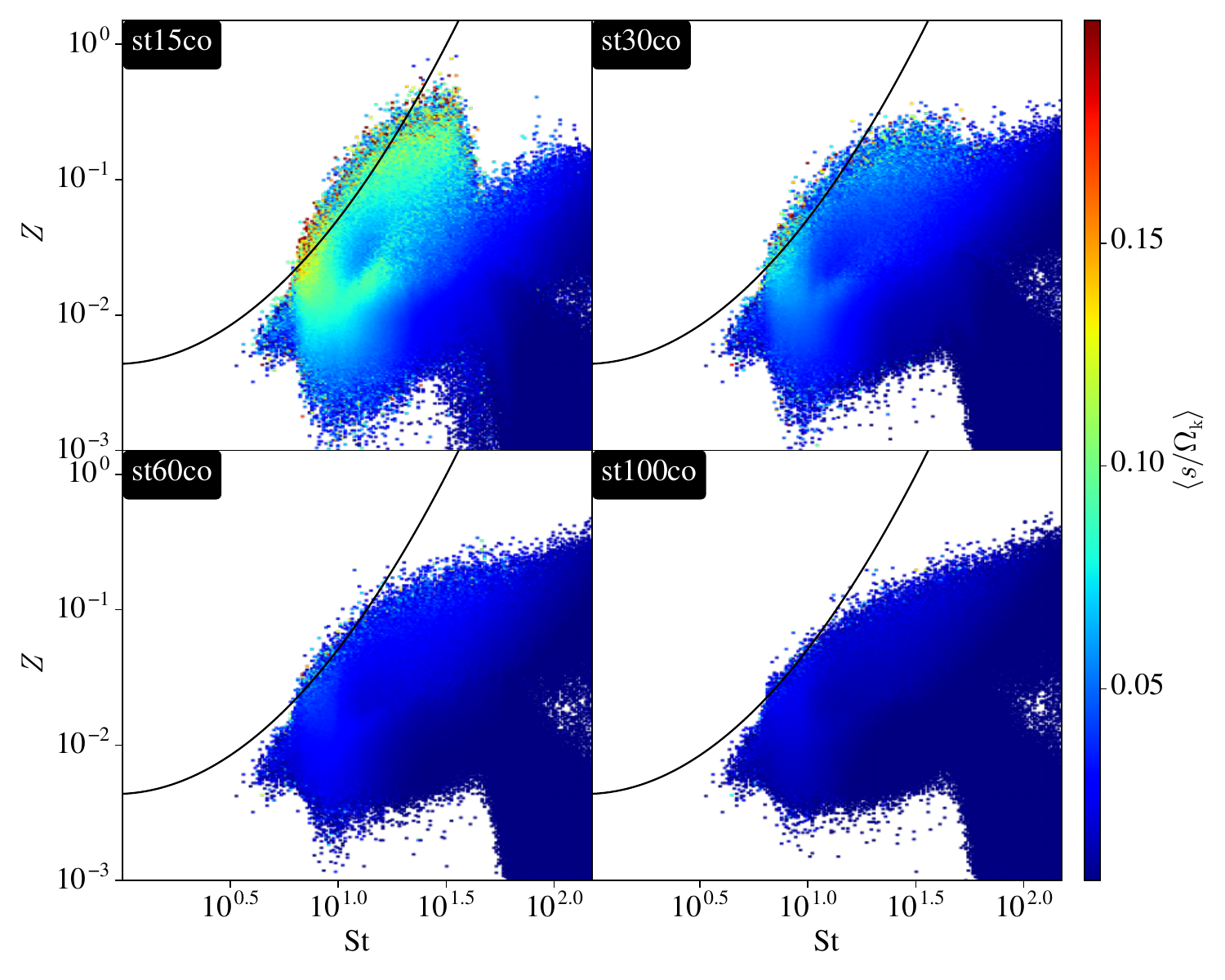}
    \caption{Pre-flyby}
    \end{subfigure}
    %\vspace{0.3em}
    \hfill
    \begin{subfigure}[t]{0.97\columnwidth}
    \centering
    \includegraphics[width=\linewidth]{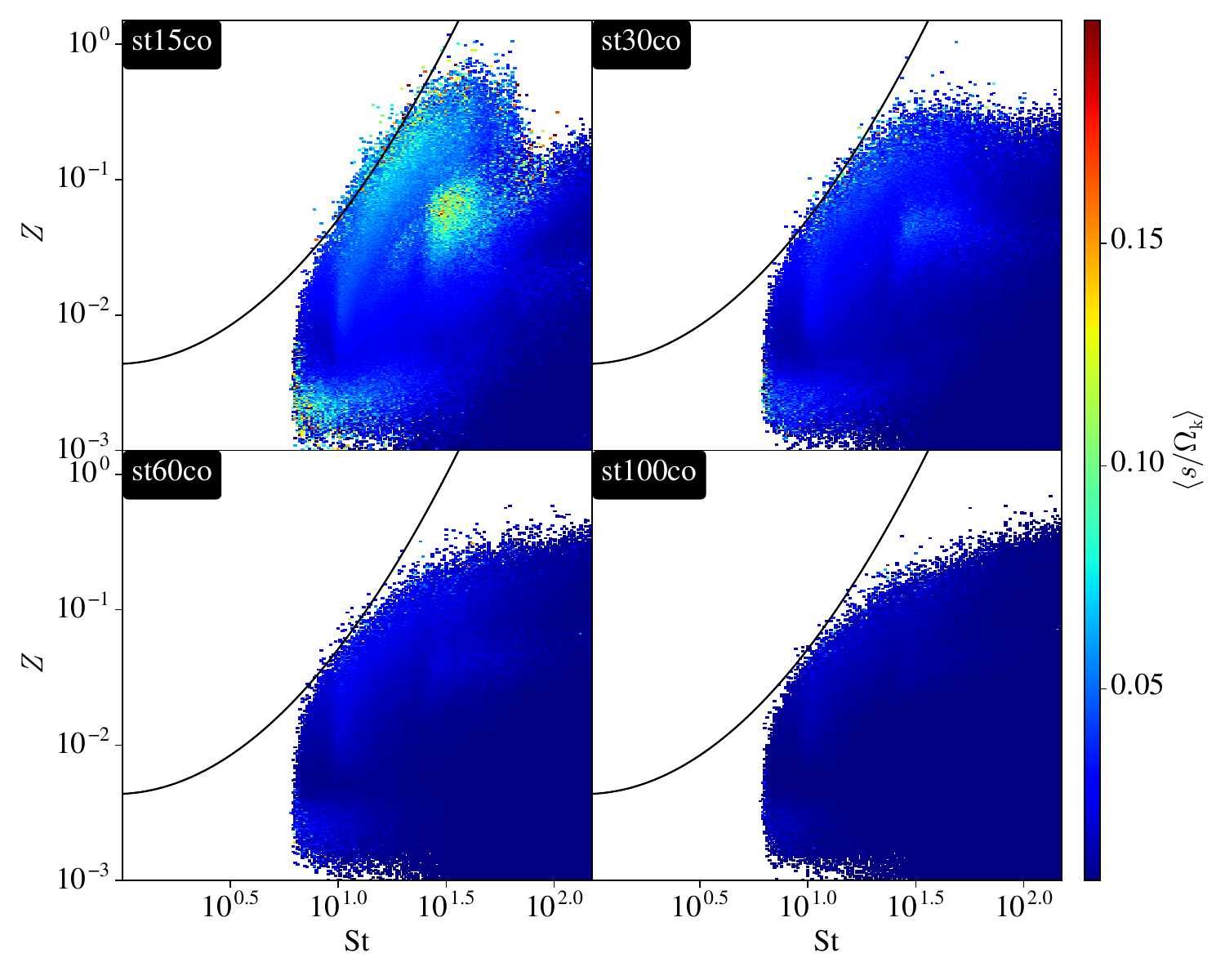}
    \caption{Suppression Phase}
    \end{subfigure}
    %\vspace{0.3em}
    \hfill
    \begin{subfigure}[t]{0.97\columnwidth}
    \centering
    \includegraphics[width=\linewidth]{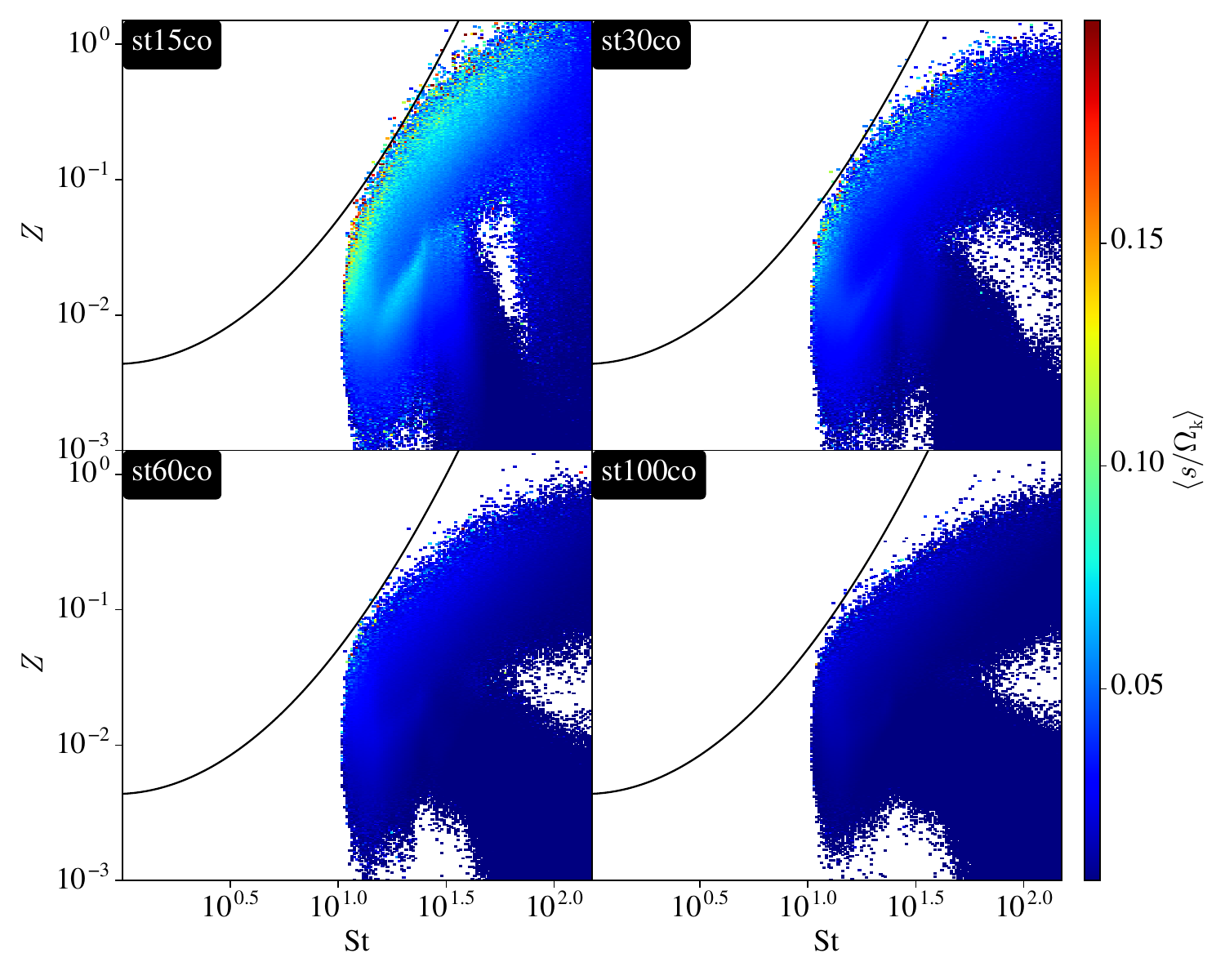}
    \caption{Recovery Phase}
    \end{subfigure}
    \caption{
    The time-averaged streaming instability growth rate as a function of the solid abundance ($Z$) and Stokes number (St) for all low-mass coplanar flyby simulations. The panels show the growth rate averaged over three distinct time intervals: pre-flyby (top), suppression (middle), and recovery (bottom). The black curve indicates the clumping threshold from \citet{Lim2025}.}
    \label{fig:MetallicityStokesDiagram}
\end{figure}

\begin{figure}
    \centering
    \hfill
    \begin{subfigure}[t]{0.97\columnwidth}
    \centering
    \includegraphics[width=\linewidth]{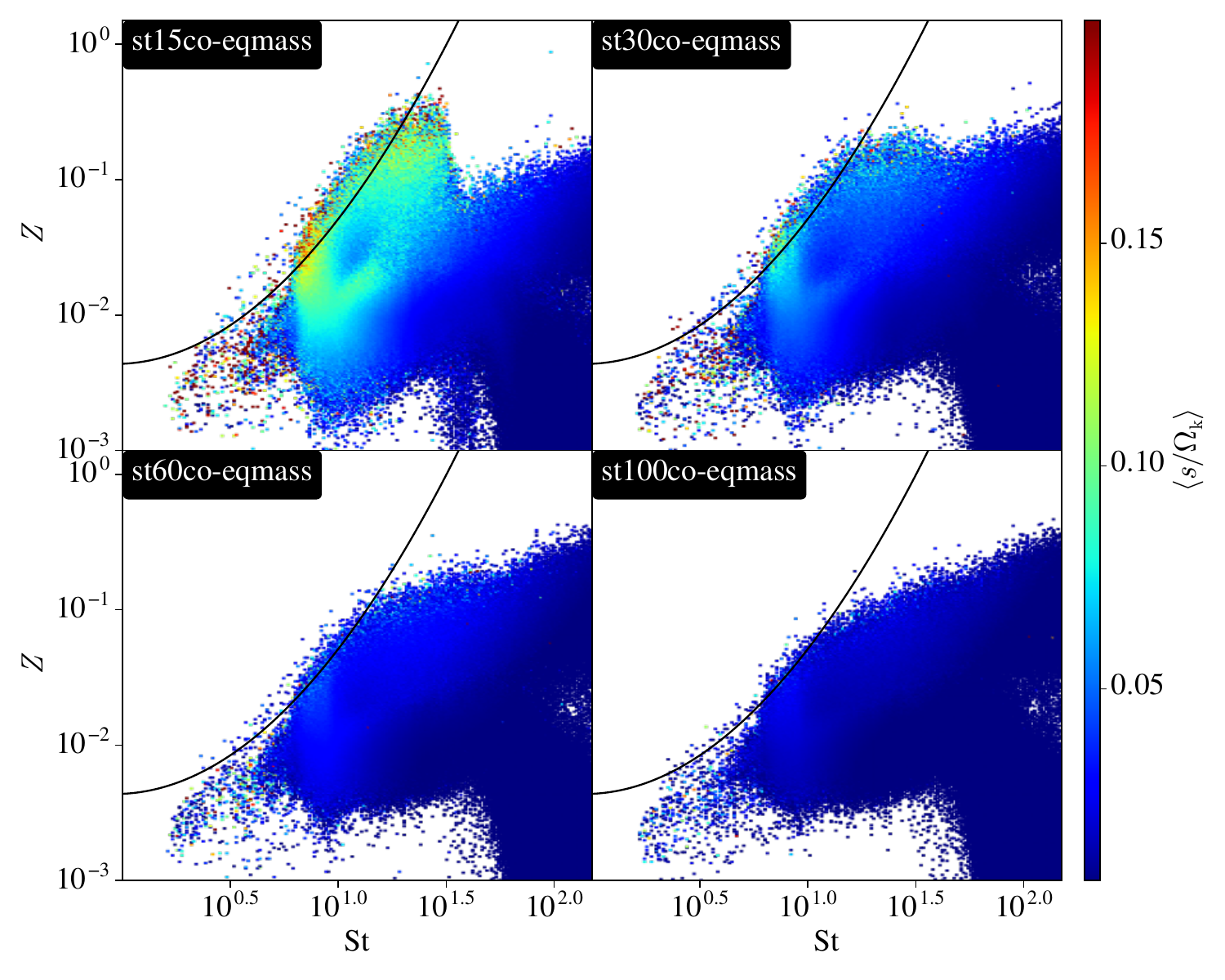}
    \caption{Pre-flyby}
    \end{subfigure}
    %\vspace{0.3em}
    \hfill
    \begin{subfigure}[t]{0.97\columnwidth}
    \centering
    \includegraphics[width=\linewidth]{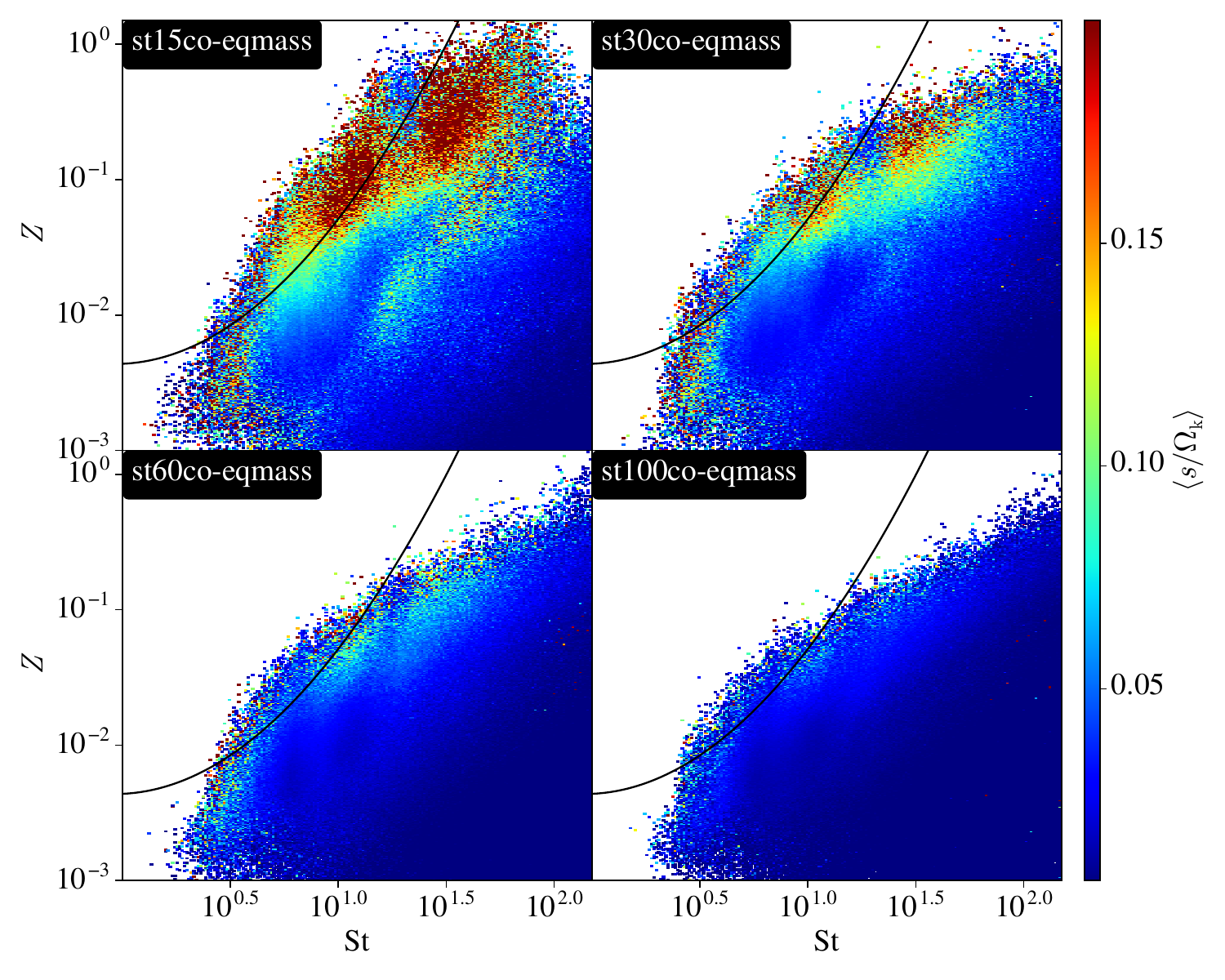}
    \caption{Suppression Phase}
    \end{subfigure}
    %\vspace{0.3em}
    \hfill
    \begin{subfigure}[t]{0.97\columnwidth}
    \centering
    \includegraphics[width=\linewidth]{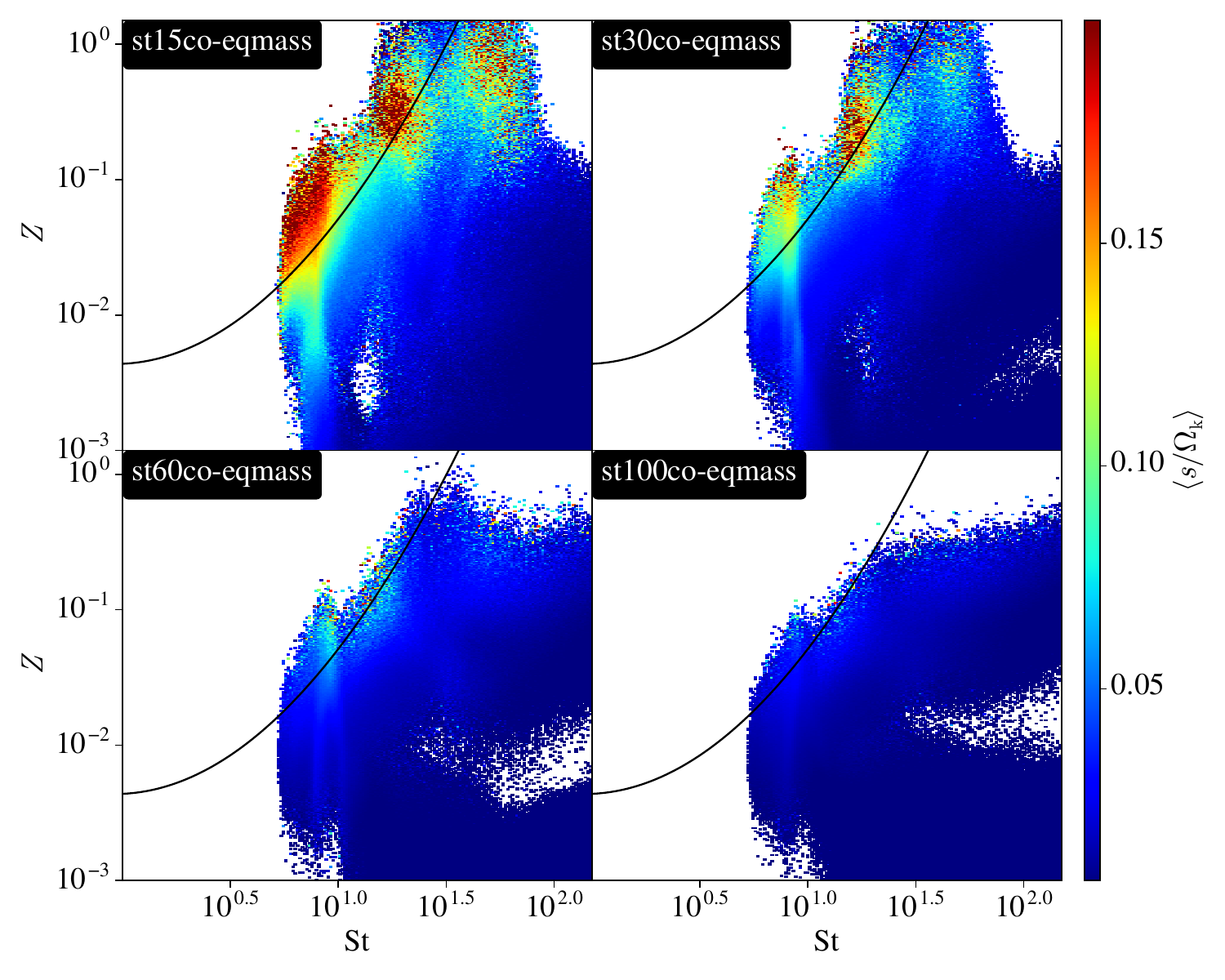}
    \caption{Recovery Phase}
    \end{subfigure}
    \caption{Same as Fig.~\ref{fig:MetallicityStokesDiagram}, but for all equal-mass flyby simulations.}
    \label{fig:MetallicityStokesDiagram_eq}
\end{figure}

We now turn our attention to investigating how flyby encounters may influence dust concentration and planetesimal formation under the framework of streaming instability, following the method described in Section \ref{sschap:MethodDustGrowth}. 

\fref{fig:Faceon_growthrate} compares the spatial distribution of the growth rate of the linear streaming instability before (top panel) and after (bottom panel) a stellar flyby in the simulation of st15co. Prior to the encounter, the growth rate is distributed inhomogeneously in small-scale, flocculent structures. After the interaction, a set of four prominent spiral arms emerges—two in the gas and two in the dust. The growth rate is significantly enhanced along the dust spirals themselves, while  the enhancement due to gas spirals primarily occurs on the radially outward side  (i.e., the leading edge) of the gaseous spiral arms. Conversely, the growth rate is suppressed in the inter-arm regions. A radial dependency is also apparent, with growth being less efficient in the inner disc compared to the outer regions. Thus, the figure indicates that while flyby-induced spirals effectively trace regions of enhanced instability growth , the rest of the disc may become more stable to the streaming instability

The time evolution of the streaming instability growth rate, averaged along each spiral arm, is shown in \fref{fig:AveragedSpiralGRSI} for systems perturbed by coplanar (left panel) and inclined (right panel) flybys. Initially, the growth rates in the dust spirals are significantly higher than in the corresponding gas spirals for both encounter types. Additionally, as the Stokes number increases, the streaming instability growth rates decrease. Moreover, for coplanar flyby–induced spirals, the growth rate averaged along each spiral arm exceeds the pre-flyby average for all four spirals (two gas and two dust) during the early phase. In inclined configurations, only the dust spirals and the gas spirals at st15in rise above the pre-flyby average, and the latter do so only briefly. A comparison between the panels reveals that the coplanar encounter induces higher streaming instability growth rates within the spirals than the inclined encounter. 

\fref{fig:AzimutahllyAveragedGrowthRate} illustrates the evolution of the azimuthally-averaged growth rate ($s/\Omega_{\rm k}$) of the streaming instability as a function of radius and time. A common trend across all simulations is a temporary suppression of the global growth rate immediately following the flyby's periastron passage, which is then followed by a recovery. While the global average is suppressed, the local growth rate is concurrently enhanced within the newly formed spiral arms (see Figure \ref{fig:Faceon_growthrate}). The nature of the recovery depends strongly on the flyby parameters. In the low-mass encounters (panels a and b), the post-flyby growth rate recovers but remains below the initial, pre-flyby level. For these cases, the growth rate consistently decreases as the dust's Stokes number increases. In stark contrast, the equal-mass coplanar flyby (panel c) induces a much more dramatic response, causing the growth rate to surge to values significantly exceeding the pre-encounter state. This result indicates that a sufficiently massive flyby can substantially boost the streaming instability in the disc. 

Based on the evolution shown in Fig.~\ref{fig:AzimutahllyAveragedGrowthRate}, we define three distinct stages:
\begin{enumerate}
    \item \textbf{Pre-flyby}: The period prior to the perturber's periastron passage, representing the disc's initial state ($t = 0 - t_\mathrm{p}$).
    \item \textbf{Suppression Phase}: The interval immediately following periastron, during which the global growth rate drops and remains low before a subsequent recovery begins ($t = 1.1 t_\mathrm{p} - 1.45 t_\mathrm{p}$).
    \item \textbf{Recovery Phase}: The period characterized by a rapid rise in the global growth rate, which eventually levels off. Prominent structures, such as the ring seen in the equal-mass coplanar case, can form during this stage ($t = 2 t_\mathrm{p} - 4 t_\mathrm{p}$).
\end{enumerate}

Using the coarse cylindrical grid described in Section~\ref{chap:analysis-method}, we distribute the measurements from the cells in the $(Z, \mathrm{St})$ space, and then time average the growth rate of the streaming instability in this space for each of the three phases, arriving at Figure~\ref{fig:MetallicityStokesDiagram} and Figure \ref{fig:MetallicityStokesDiagram_eq}.  Figure~\ref{fig:MetallicityStokesDiagram} is for low-mass coplanar flyby simulations. Prior to the encounter (top panel), a significant region of the disc has $Z$ above the critical clumping threshold from Eq.~\ref{eq::z_crit}, indicating conditions favorable for dust clumping, and this region also shows high average growth rate. However, immediately following periastron (middle panel), both $Z$ and the growth rate are substantially reduced, falling below the critical curve and suggesting that the flyby initially halts the clumping process. In the subsequent recovery phase (bottom panel), the growth rate increases but the solid abundance fails to reach its pre-flyby magnitude, remaining below the critical threshold and implying that dust clumping remains restricted long after the encounter. Across all three phases, the growth rate consistently decreases with increasing grain size.

Figure \ref{fig:MetallicityStokesDiagram_eq} presents the same analysis as Figure \ref{fig:MetallicityStokesDiagram}, but for the equal-mass flyby simulations. The pre-flyby state (top panel) is similar to the low-mass case, with a region of high growth rate above the critical clumping threshold from Eq.~\ref{eq::z_crit}. However, in stark contrast to the low-mass companion, the post-periastron evolution for the equal-mass flyby shows a dramatic surge in both the local solid abundance and the growth rate  of the streaming instability (middle and bottom panels). The region of enhanced abundance expands substantially, rising well above the critical curve to values that exceed even the initial pre-flyby state. This result demonstrates that an equal-mass flyby encounter can strongly promote the conditions required for dust clumping, rather than suppressing them. 

From our simulations with initial averaged Stokes numbers of $\mathrm{St} \sim$ 15, 30, 60, and 100, we find that the SI growth rate becomes larger for smaller Stokes numbers within this sampled range. Although we do not simulate particles with $\mathrm{St} \lesssim$ 15, it is reasonable to expect, by extrapolation, that the growth rate would further increase toward $\mathrm{St} \sim$ 1. This expectation arises because dust particles with Stokes numbers close to unity drift most efficiently and accumulate in pressure maxima \citep{Gonzalez2017}, a behaviour also noted in Section~\ref{sschap:DiscMorphology}.

\section{Discussion}
\label{sec::Caveats}
Our simulations present a simplified model of the dust-gas dynamics in protoplanetary discs under a stellar flyby, and several important processes are not included, which could influence the results. First, coagulation and fragmentation of dust particles \citep{Birnstiel2024} are not incorporated into the simulations.  Furthermore, the primary star at the centre of the circumprimary disc is modelled as a standard sink particle. This simplification neglects the effects of radiation pressure and magnetic fields on the discs. Stellar irradiation can generate a positive vertical temperature gradient \citep{Chiang1997}, thereby modifying the thermal structure. Magnetic field processes, in particular the Magneto-Shear Instability (MSI) that may promote dust concentration \citep{Lin2022}, are likewise excluded from our approach. Additionally, the interaction between the flyby and the protoplanetary disc can lead to disc heating through compression and shocks, as demonstrated by \cite{Lodato2007a}. This heating could significantly alter the disc morphology and its dynamics. Our simulations, however, do not account for such heating processes, and as a result, thermal effects, including those arising from temperature gradients as discussed by \cite{Lee2015} and \cite{Juhasz2018}, are not incorporated into our model. These limitations highlight the need for future studies to explore additional physical effects, such as self-gravity and radiative feedback, to better capture the complexity of dust evolution in flyby encounters. A natural extension of this work would be to investigate dust coagulation and fragmentation in a system undergoing stellar flybys. This could be carried out by applying the model proposed by \cite{Vericel2022} to examine the dust size distribution and collisional outcomes across various types of flyby events.

During the review of this work, \cite{Prasad2025} examined substructures in flyby-perturbed discs within a low Stokes number regime, assessing whether conditions might trigger the planetesimal formation based on the clumping criteria of \cite{Li2021}. While their approach identified regions that satisfy necessary conditions for clumping, it did not quantify the growth rates of the instability. Our study provides a different perspectives by calculating the streaming instability growth rates from the simulations using the framework of \cite{Chen2018}. This requires evaluating gas and dust properties at the disc midplane via SPH interpolation, where the streaming instability develops most efficiently. Although we operate in a larger Stokes number regime, both studies reach the same conclusion: flyby encounters may promote the streaming instability.

Our analysis assumes a gas surface density profile exponent of $p=1.5$, consistent with the Minimum Mass Solar Nebula (MMSN) model. In contrast, previous hydrodynamic simulations of flyby–disc interactions used a shallower profile of $p=1$ to match observed disc profiles \citep{Cuello2019,Cuello2020}. Steeper profiles enhance radial drift but also deplete the mass in the outer regions of the disc \citep{Laibe2014}. Conversely, shallower profiles reduce radial drift, which affects the growth rate of the streaming instability. In principle, the streaming instability requires a moderate drift velocity; if the drift is too rapid, dust cannot be concentrated, and if it is too slow, there is insufficient energy to trigger the mechanism \citep{Brauer2008}. The surface density profile also affects the back-reaction of dust on gas, a crucial component of the streaming instability \citep{Gonzalez2017,Magnan2024}. A detailed analysis of how $p$ affects streaming instability growth rates in protoplanetary discs encountered by flyby companions is beyond the scope of this paper. 

Particles with Stokes numbers $\mathrm{St}\sim1$ are known to drift efficiently and concentrate in pressure maxima, making them highly susceptible to the streaming instability. Our simulations, which model particles with initial Stokes numbers as low as 15, reveal a clear trend: the streaming instability growth rate increases as the Stokes number decreases (see Fig.~\ref{fig:MetallicityStokesDiagram}). This result suggests that particles in the $\mathrm{St}\sim 1$ regime would be even more unstable to the streaming instability after flyby encounters. This has direct observational implications, as facilities like ALMA and the VLA are most sensitive to dust populations corresponding to these Stokes numbers in protoplanetary discs.

\section{conclusion}
\label{sec::conclusions}

In this work, we used three-dimensional SPH simulations to explore the dynamical response of gas and dust discs to prograde stellar flybys, varying the perturber's mass, inclination, and the dust's Stokes number. Our simulations demonstrate that flybys robustly trigger the formation of spiral arms in both components, though their evolution depends strongly on the encounter's mass. For low-mass flybys, gaseous arms tend to dissipate while dust spirals persist and become more tightly wound over time, leading to an increasing offset and different pattern speeds between the two. In contrast, an equal-mass flyby induces such tightly wound dust spirals that they merge into a distinct, eccentric ring-like structure that migrates inwards.  While all flybys excite the disc's eccentricity, particularly in the outer regions, the key implication is that the induced structures—both transient spirals and persistent rings—act as effective dust traps. 

The impact of a stellar flyby on the conditions for dust clumping is critically dependent on the perturber's mass. A low-mass flyby, despite creating local enhancements, ultimately suppresses the dust concentrations to a level below the critical threshold for clumping, even after the disc recovers. In contrast, an equal-mass flyby powerfully promotes local dust concentrations, especially along the dust spiral arms; following a brief disruption, it triggers a surge in the local solid abundance that significantly exceeds both the pre-flyby state and the critical clumping threshold. Thus, while low-mass encounters may inhibit planetesimal formation via this mechanism, massive flybys can establish exceptionally favorable conditions for rapid dust clumping. However, high dust concentrations do not guarantee coagulation, as high-velocity collisions can instead lead to fragmentation, a possibility that warrants future study.

\section*{Acknowledgements}
The authors thank the referee for a prompt and constructive review. This work is supported by the Taiwan Ministry of Science and Education and the ASIAA Summer Student Program. JLS acknowledges funding from the Dodge Family Prize Fellowship in Astrophysics. JLS is also supported by the Vice President of Research and Partnerships of the University of Oklahoma and the Data Institute for Societal Challenges. 
CCY acknowledges the support from NASA via the Emerging Worlds program (\#80NSSC23K0653), the Astrophysics Theory Program (grant \#80NSSC24K0133), and the Theoretical and Computational Astrophysical Networks (grant \#80NSSC21K0497).
NC acknowledges funding from the European Research Council (ERC) under the European Union Horizon Europe programme (grant agreement No. 101042275, project Stellar-MADE).

%%%%%%%%%%%%%%%%%%%%%%%%%%%%%%%%%%%%%%%%%%%%%%%%%%
\section*{Data Availability}
The data supporting the plots within this article are available on reasonable request to the corresponding author. The analysis routine is available at \url{https://github.com/weishansu011017/PhantomRevealer.jl}. A public version of the {\sc phantom}, and {\sc splash} codes are available at \url{https://github.com/danieljprice/phantom}, and \url{http://users.monash.edu.au/~dprice/splash/download.html}, respectively.



\bibliographystyle{mnras}
\bibliography{ref.bib} % if your bibtex file is called example.bib

%\begin{thebibliography}{99}
%\end{thebibliography}

%%%%%%%%%%%%%%%%%%%%%%%%%%%%%%%%%%%%%%%%%%%%%%%%%%

%%%%%%%%%%%%%%%%% APPENDICES %%%%%%%%%%%%%%%%%%%%%

\appendix
\section{Kernel interpolation in 3D SPH simulation}\label{app:KernelInterpolation}
% -*Column density distribution*
\begin{figure}
    \centering
    \includegraphics[width=\columnwidth]{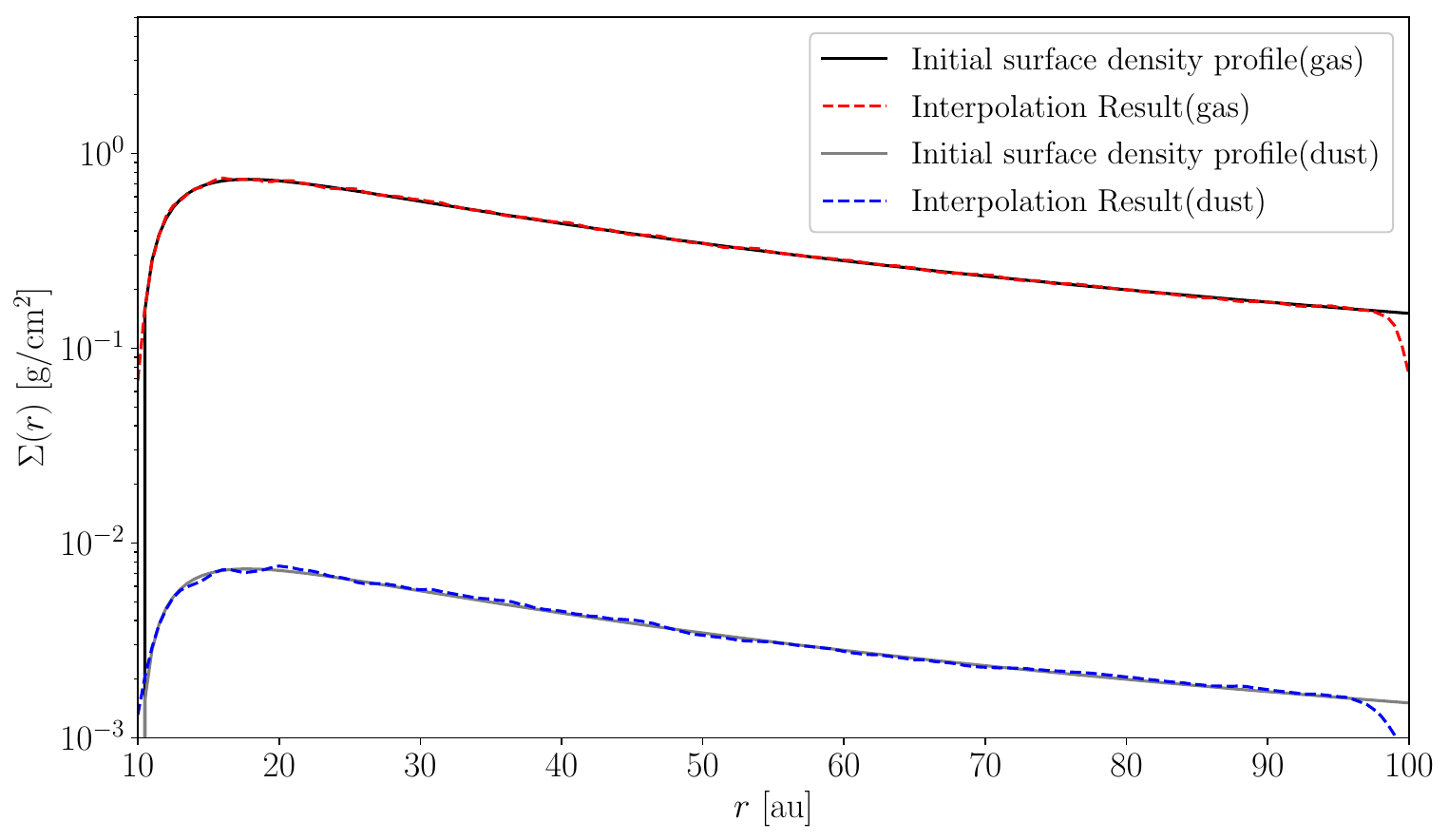}
    \caption{A comparison of initial column density profile between theoretical profile and interpolation by post-processing of the initial data. The solid line represent the column density profile that is used for constructing the disc in \textsc{phantom}, and the dashed line is the profile from interpolation.}
    \label{fig:InitialSurfaceDensity}
\end{figure}
Following the SPH kernel interpolation proposed in \cite{Price2012a}, which adopts a particularly concise and elegant formulation, written as 
\begin{equation}
    \mathbf{A}(\mathbf{r}) = \frac{\sum_b^{N_{\mathrm{neigh}}} \frac{m_b\mathbf{A}_b}{\rho_b}W(\mathbf{r} - \mathbf{r}_b; h)}{\sum_b^{N_{\mathrm{neigh}}} \frac{m_b}{\rho_b}W(\mathbf{r} - \mathbf{r}_b; h)}
\end{equation}
where $\mathbf{A}$ is an arbitrary continuous physical quantities that would be interpolated from particles, $m_b$ is the mass of particles, and $\rho_b$ is the local fluid density at particles position $\mathbf{r}_b$. $h$ is the smoothed radius at the interpolation point $\mathbf{r}$. $W$ is the kernel function for SPH, which has different truncated radius $R_{\mathrm{truncated}}$ such that $W(r> R_{\mathrm{truncated}})= 0$. Hereafter, a sphere of radius $R_{\mathrm{truncated}}$ centred at the interpolation point $\mathbf{r}$ will be referred to as a truncated sphere. $N_{\mathrm{neigh}}$ is the neighbourhood around the interpolation point $\mathrm{r}$, which includes those particles located insides the truncated sphere. In our implementation, $M_6$ quintic B-spline function is chosen for interpolation, with $R_{\mathrm{truncated}} = 3h$.
\begin{equation}
    W_{M_{6}}(\zeta) = N
    \begin{cases}
        (3-\zeta)^5 - 6(2-\zeta)^5 + 15(1-\zeta)^5 & 0\leq \zeta <1\\
        (3-\zeta)^5 - 6(2-\zeta)^5 & 1\leq \zeta <2\\
        (3-\zeta)^5 & 2\leq \zeta <3\\
        0 & \zeta\geq 3
    \end{cases}
\end{equation}
where $\zeta = \left| \mbf{r} - \mbf{r}^\prime \right| /h$, $N$ is a normalized constant with $1/120\pi h^{3}$ in 3D, which is given in \cite{Price2012a}. We adopt the line integration interpolation as the same as {\sc splash} \citep{Price2007} along the z-axis to calculate the integration value. 
\par

We implement the kernel interpolation with {\sc julia}, a high-level, general-purpose dynamic programming language commonly used in scientific computing and visualization \citep{Julia-2017}, to process and compute the physical quantities of interest from the simulation. According to the kernel function, the influence of particles is only bounded within a radius of $3h$. Thus, an algorithm that can search its nearest neighbours at a specific position with a given radius is required. To reduce the time required for searching for nearest neighbours, we employ a k-dimensional tree, known as a KD-tree, which is an efficient method for locating particles near a given position in our algorithm. The KD-tree is a data structure designed for storing information that can be retrieved through associative searches \citep{Bentley1975}. It reduces the searching time complexity from $\mathcal{O}(N)$ to $\mathcal{O}(\log N)$ \citep{Skrodzki2019}. In our implementation, the KD-tree is provided by the Julia module \textit{Neighborhood.jl\footnote{\url{https://github.com/JuliaNeighbors/Neighborhood.jl}} }.

To check our implementation of interpolation, we compared the column density profile from kernel interpolation and Equation \ref{eq:SurfaceDensityProfile} in Figure \ref{fig:InitialSurfaceDensity} at $\mathrm{t}=0 \,\mathrm{yr}$, and it closely recovered our initial profile Equation \ref{eq:SurfaceDensityProfile}.\par

\section{Spiral detection}\label{app:SpiralDetection}
\begin{figure}
\centering
\begin{tikzpicture}[node distance=1.5cm]
    \node (SPHParticlesData) [startstop] {SPH particles data};

    \node (LOSitp) [process, below of=SPHParticlesData, align=center] {Line integration interpolation\\in polar grid \citep{Price2007}};

    \node (ColumnDensity) [midprod, below of=LOSitp] {$\Sigma(r,\phi)$};

    \node (RidgeDetection) [process, below of=ColumnDensity, align=center] {Ridge detection\\with automatic scale selection \citep{Lindeberg1998}};

    \node (RidgePoints) [midprod, below of=RidgeDetection, align=center] {Detected\\"Ridge"\\points};

    \node (StrengthRidgePoints) [midprod, below of=RidgeDetection, xshift=-2.5cm]{$\mathcal{N}_{\gamma-\mathrm{norm}}L$};

    \node (Width2RidgePoints) [midprod, below of=RidgeDetection, xshift=2.5cm] {$\tau$};

    \node (HoughTransform) [process, below of =RidgePoints, align=center] {Strength-weighted Hough Transform \citep{Hough1962}\\in $(\ln a, k)$-space};

    \node (Accumulator) [midprod, below of=HoughTransform] {Hough accumulator};
    
    \node (ExtractPeak) [process, below of=Accumulator, align=center] {Extract accumulator peaks\\(Potential spiral pattern)};

    \node (SpiralCandidates) [midprod, below of=ExtractPeak] {Spiral candidates};

    \node (BeamSearch) [process, below of=SpiralCandidates, align=center] {Beam search \citep{Lowerre1976, Graves2012}\\with strength-weighted coverage scoring and structural penalty};

    \node (Spirala) [startstop, below of=BeamSearch, align=center, , xshift=-2.5cm] {$\lbrace a_i\rbrace$ parameters\\for detected\\ logarithmic spirals};
    
    \node (SpiralCoverage) [startstop, below of=BeamSearch, align=center] {Detected\\"Spiral"\\points};
    
    \node (Spiralk) [startstop, below of=BeamSearch, align=center, xshift=2.5cm] {$\lbrace k_i\rbrace$ parameters\\for detected\\ logarithmic spirals};

    % Arrow
    \draw [arrow] (SPHParticlesData) -- (LOSitp);
    \draw [arrow] (LOSitp) -- (ColumnDensity);
    \draw [arrow] (ColumnDensity) -- (RidgeDetection);
    \draw [arrow] (RidgeDetection) -- (RidgePoints);
    \draw [arrow] (RidgeDetection) -- (StrengthRidgePoints);
    \draw [arrow] (RidgeDetection) -- (Width2RidgePoints);
    \draw [arrow] (RidgePoints) -- (HoughTransform);
    \draw [arrow] (StrengthRidgePoints.west) -- ++(-0.5,0) -- ++(0,-1.5) --(HoughTransform.west);
    \draw [arrow] (HoughTransform) -- (Accumulator);
    \draw [arrow] (Accumulator) -- (ExtractPeak);
    \draw [arrow] (ExtractPeak) -- (SpiralCandidates);
    \draw [arrow] (SpiralCandidates) -- (BeamSearch);
    \draw [arrow] (StrengthRidgePoints.west) -- ++(-0.5,0) -- ++(0,-7.5) --(BeamSearch.west);
    \draw [arrow] (Width2RidgePoints.east) -- ++(0.5,0) -- ++(0, -7.5) -- (BeamSearch.east);
    \draw [arrow] (BeamSearch) -- (Spirala);
    \draw [arrow] (BeamSearch) -- (Spiralk);
    \draw [arrow] (BeamSearch) -- (SpiralCoverage);

    % Other relation
    \draw[dashed] (RidgePoints) -- (StrengthRidgePoints);
    \draw[dashed] (RidgePoints) -- (Width2RidgePoints);

    \draw[dashed] (SpiralCoverage) -- (Spirala);
    \draw[dashed] (SpiralCoverage) -- (Spiralk);

\end{tikzpicture}
\caption{Working process for spiral detection. $\mathcal{N}_{\gamma-\mathrm{norm}}L$ is the ridge strength for each “detected points”, which indicates how confident the detection is to be true. $\tau$ is the scale of detected ridges, which inform the approximate width of ridge at the ridge point $p_i$ with width $\sqrt{\tau_i}$.}
\label{fig:SpiralDetectionFlowChart}
\end{figure}
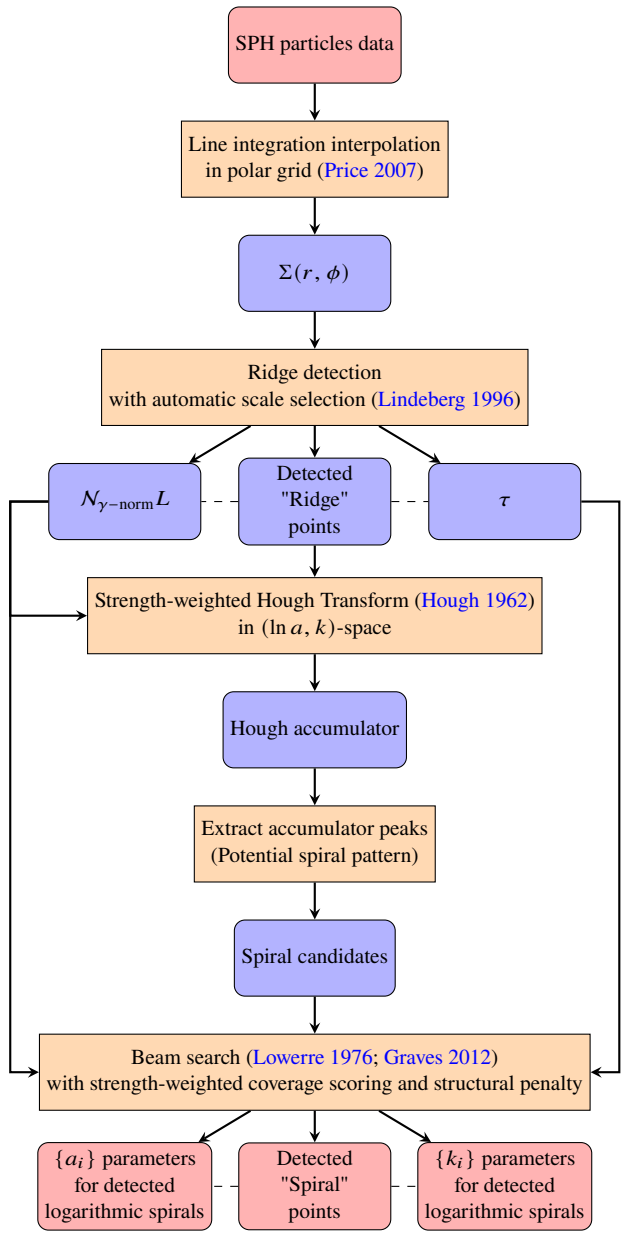
\begin{figure}
    \centering
    \includegraphics[width=\columnwidth]{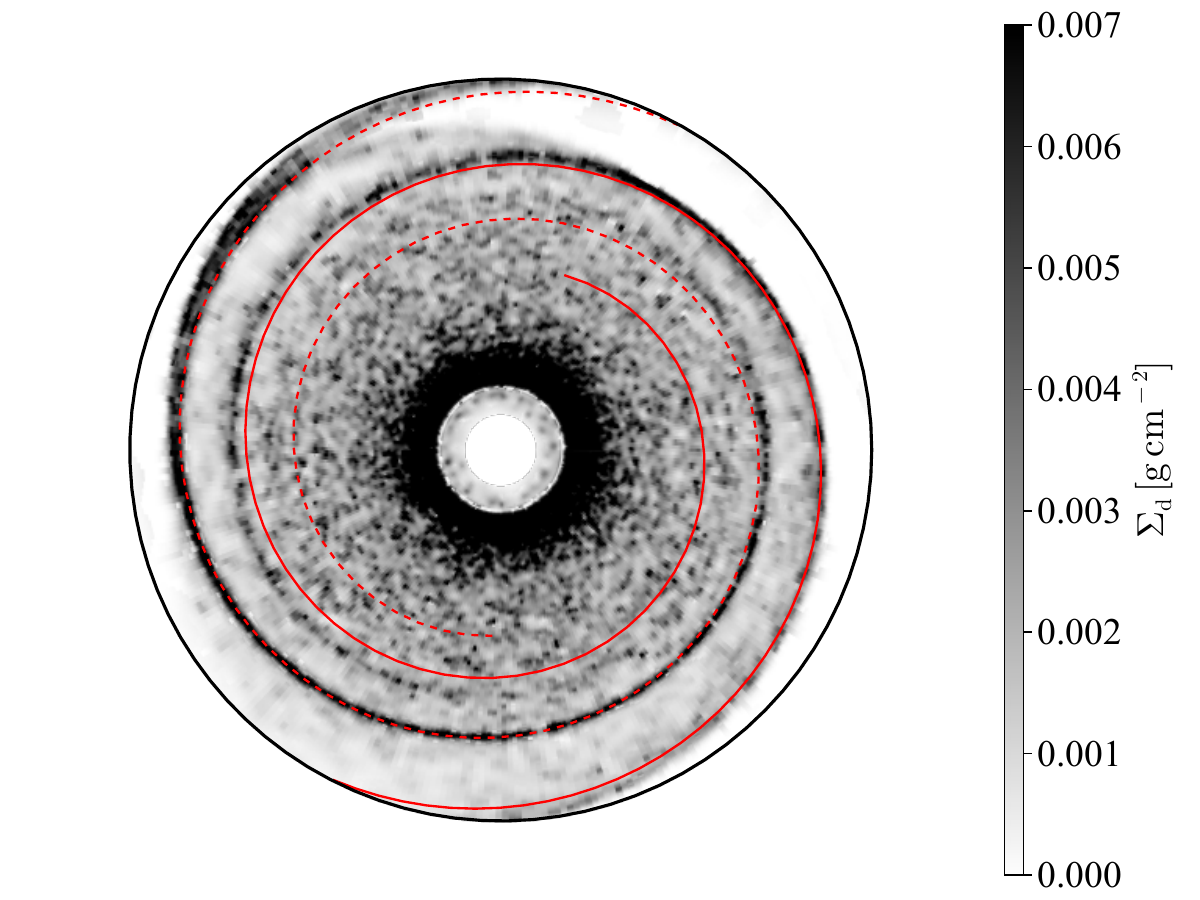}
    \caption{An example of the spiral detection method applied to a dusty protoplanetary disc. The gray scale indicates the dust column density. The two detected spiral arms are overplotted as lines: arm 1 (solid) and arm 2 (dashed).}
    \label{fig:SpiralDetection}
\end{figure}
We describe in this section the spiral detection technique we use to detect spirals in our simulations. The flow chart of full detection process is shown in \fref{fig:SpiralDetectionFlowChart}. After processing the column density distribution in a polar grid, labelled as $\Sigma(r,\phi)$, the detection process can roughly break down into three parts: ridge detection, Hough transform, and the Beam search. 
\subsection{Ridge detection}\label{sapp:RidgeDetection}
We adopted the ridge detection with automatic scale selection proposed by \cite{Lindeberg1998}. As the first step, we convert $\Sigma$ into the scale-space representation:
\begin{equation}
    L(x_{\mathrm{pix}}, y_{\mathrm{pix}};\tau) = \mathcal{G}(x_{\mathrm{pix}}, y_{\mathrm{pix}};\tau) *\Sigma(x_{\mathrm{pix}}, y_{\mathrm{pix}})
\end{equation}
Since ridge detection requires a well-defined metric for distances and gradients, we evaluated both physical and grid-based pixel coordinates. The physical metric was found to be unsuitable because its inherent radial anisotropy led to unstable results. Consequently, we discarded the physical metric and conducted the ridge detection in pixel coordinates. This approach uses a uniform metric defined by our grid spacing ($\Delta r=0.47\, \rm au$, $\Delta \phi=0.017\, \rm rad$) and is independent of the physical length scales. For simplicity, $(x_{\mathrm{pix}}, y_{\mathrm{pix}})$ will be denoted as $(x,y)$ in this subsection. $\tau$ is the “scale” of the ridge detection, which represent the approximate width-square of the ridge in pixel units. $\mathcal{G}(x, y;\tau)$ is a 2D Gaussian distribution. 
\begin{equation}
    \mathcal{G}(x, y;\tau) =\frac{1}{2\pi \tau}e^{-(x^2 + y^2)/2\tau}
\end{equation}
The scale-space derivatives are defined by
\begin{equation}
    L_{x^{\alpha}y^{\beta}} = \partial_{x^{\alpha}y^{\beta}}L = \mathcal{G}_{x^{\alpha}y^{\beta}} * \Sigma
\end{equation}
where $(\alpha, \beta)$ represent the order of differentiation. These operations are implemented by Fast Fourier Transform (FFT) based on the {\sc julia} package \textit{FFTW.jl \footnote{\url{https://github.com/JuliaMath/FFTW.jl}}}.\\
Next, for each pixel $(x_0,y_0)$ in $\Sigma$, transforming the coordinate into $(p,q)$-space such that $L_{pq} =0$. The rotation operator is performed in \cite{Lindeberg1998}. Finally, those points which satisfied
\begin{equation}
    \begin{cases}
        L_p = 0 \\
        L_{pp} <0 \\
        |L_{pp}| \geq |L_{qq}|
    \end{cases}
    \mathrm{or}
    \begin{cases}
        L_q = 0 \\
        L_{qq} <0 \\
        |L_{qq}| \geq |L_{pp}|
    \end{cases}
\end{equation}
would be labelled as “Ridge points”. Note that detection at different scales is sensitive to ridges of different thicknesses. Therefore, we adopt this detection within a scale interval $5 <\tau <22$. To determine the most appropriate ridge width across different detection scales, we estimate the ridge strength.
\begin{equation}
    \mathcal{N}_{\gamma-\mathrm{norm}}L = \tau^{4\gamma}(L_{xx} + L_{yy})^2((L_{xx}-L_{yy})^2+4L_{xy}^2)
\end{equation}
where $\gamma$ is a parameter for the normalized derivative \citep{Lindeberg1998}, and is set to 0.75. \\
All detected ridge points are aggregated into a binary map, along with their corresponding ridge strength and scale. These three pieces of information form the foundation for the subsequent procedures.
\subsection{Strength-weighted Hough Transform}
After extracting ridge points, the next step is to fit spiral patterns from the resulting point set. However, conventional gradient descent methods are not suitable in this scenario, as the data may contain multiple spirals with no known parametric equations to describe them accurately. Moreover, the spiral parameters are highly sensitive; even small variations in their values can produce large changes in the location and shape of the fitted spiral. To address these issues, we adopt a modified Hough transform to detect spiral structures. Hough transform \citep{Hough1962} is a technique for line detection. It maps each point in the input space to a line in the parameter space corresponding to the slope-intercept form $y = mx + b$  where the accumulation of votes indicates the most probable set of parameters $(m,b)$. In our scenario, a logarithmic spiral in the  $(\ln r, \phi)$ space can be written as follows:
\begin{equation} \label{eq:logspiral}
    \ln r = k\phi + \ln a
\end{equation}
where $(k,a)$ is the parameters of  logarithmic spiral, with the conditions $a>0$ and $k\neq 0$. Equation \ref{eq:logspiral} indicates that the  logarithmic spiral is a straight line in the $(\ln r, \phi)$ space. 

In our implementation, we slice the $\ln a$ axis into 450 segments within the interval $3.4 < \ln a < 5.7$, the $k$ axis into 80 segments within the interval $-0.25 < k < -0.06$. Besides, the voting process is weighted by the ridge strength at each point, allowing stronger ridge responses to contribute more significantly in the Hough space. 

Importantly, spirals in our simulation have thickness, which leads to multiple peaks existing in the accumulator for a single spiral. Moreover, the point set contains multiple spirals, resulting in multiple “peak cluster” in the accumulator. Worse still, in the polar grid, a single peak cluster may be split into multiple parts due to the cyclic nature of $\phi$, causing nearly impossible to separate the peak clusters for each spiral directly. To address these issues, we collect all the peaks in the accumulator and match them to the most likely spiral combination in the subsequent step.

\subsection{Beam search with strength-weighted coverage scoring and structural penalty}
Given a collection of spiral candidates extracted from the Hough accumulator, the task is to select a subset that collectively explains the ridge structure with maximal strength-weighted coverage. To this end, we define a scoring function that measures the proportion of total ridge strength covered by a candidate subset, while incorporating several penalty terms to suppress undesired configurations. For each spiral candidate $(a, k)$, we construct a binary array that represents its coverage region on the image. This region is defined by a width function determined at each radius $r$ as the mean of width $\sqrt{\tau}$ along the $\phi$-direction. When testing multi-spiral configurations, multiple spirals can be simultaneously composed into the same image.
The scoring follows by
\begin{equation}
    \mathcal{S} = \mathcal{C} - \lambda_{\mathrm{angle}}\mathcal{P}_{\mathrm{angle}} -  \lambda_{\mathrm{overlap}}\mathcal{P}_{\mathrm{overlap}}
\end{equation}
where $\mathcal{S}$ is the final score of this combination. $\mathcal{C}$ is the strength-weighted coverage, which is given by
\begin{equation}
    \mathcal{C} = \frac{\sum_{\mathrm{Covered}}\mathcal{N}_{\gamma-\mathrm{norm}}L}{\sum_{\mathrm{Total}}\mathcal{N}_{\gamma-\mathrm{norm}}L }.
\end{equation}
$\mathcal{P}_{\mathrm{angle}}$ penalizes multi-spiral configurations in which the angular positions $\phi(r_{\mathrm{ref}})$ at a reference radius $r_{\mathrm{ref}}$ of the constituent spirals are not approximately evenly spaced over the full $2\pi$ range. If the angular separation closely approximates uniform spacing, the penalty smoothly decays from 1 toward 0. The corresponding $\lambda_{\mathrm{angle}}$ is user-controlled and is set to 1. $\mathcal{P}_{\mathrm{overlap}}$ penalizes the degree to which a newly proposed spiral candidate redundantly covers regions already explained by previously selected spirals. Given two binary masks—$\mathcal{M}_{\mathrm{new}}$ and $\mathcal{M}_{\mathrm{old}}$ —that indicate the coverage regions of the current and an existing spiral, respectively, the penalty is computed as
\begin{equation}
    \mathcal{P}_{\mathrm{overlap}} = \frac{|\mathcal{M}_{\mathrm{new}} \cap \mathcal{M}_{\mathrm{old}}|}{\min(|\mathcal{M}_{\mathrm{new}}|, |\mathcal{M}_{\mathrm{old}}|)}
\end{equation}
where $|\cdot|$ denotes the number of pixels with non-zero (true) values. This normalized intersection score ensures that a complete overlap—relative to the smaller of the two masks—yields a maximal penalty of 1, while disjoint coverage results in zero penalty. The corresponding $\lambda_{\mathrm{overlap}}$ is user-controlled and we adopt 1.2, as tests on our data demonstrated that this value provides the most stable detection results. Both penalties are specifically designed for multi-spiral configurations, and thus vanish when only a single spiral is selected.

To identify the optimal spiral combination, we maximize $\mathcal{S}$ by evaluating all possible subsets of spiral candidates, including both single and multiple-spiral configurations. To systematically explore possible spiral combinations, we consider a staged construction process: starting from an empty set, spiral candidates are added one by one to form partial combinations. Each stage corresponds to a specific “combination depth”, $\mathcal{D}$, defined as the number of spirals selected so far. Any combination yielding a negative score is discarded without further consideration. 

This conceptual depth allows us to organize the search space hierarchically, where deeper levels represent more complete configurations, up to the user-controlled maximum number of spirals, $N_{\mathrm{max\,spiral}}$. To avoid overfitting to increasingly complex spiral combinations, we enforce a hard cutoff: if any configuration achieves a score above 0.75, the search is immediately terminated and the corresponding result is returned. 

However, the number of possible spiral combinations grows exponentially with the number of candidates, making exhaustive evaluation computationally impractical. To address this issue, we adopt beam search to improve computational efficiency. Beam search is a greedy, breadth-limited search algorithm that explores the search space layer by layer, retaining only the top-scoring candidate combinations at each depth to limit computational cost, as originally proposed by \citet{Lowerre1976}. Our implementation follows the structure outlined in \citet{Graves2012}, adapted to our domain-specific scoring function. At each depth $\mathcal{D}$, we retain the top 20\% of candidate combinations ranked by the scoring function $\mathcal{S}$, pruning the rest to limit the search space. This is based on the assumption that low-scoring candidates are unlikely to achieve significantly higher scores even when extended with additional spiral patterns. After these processes, a set of spiral structures is extracted from the raw input image. \fref{fig:SpiralDetection} illustrates an example result of our spiral detection pipeline.

\section{Resolution study of dust spirals}\label{app:DustSpiralResolutionStudy}
\begin{figure*}
    \centering
    \includegraphics[width=2\columnwidth]{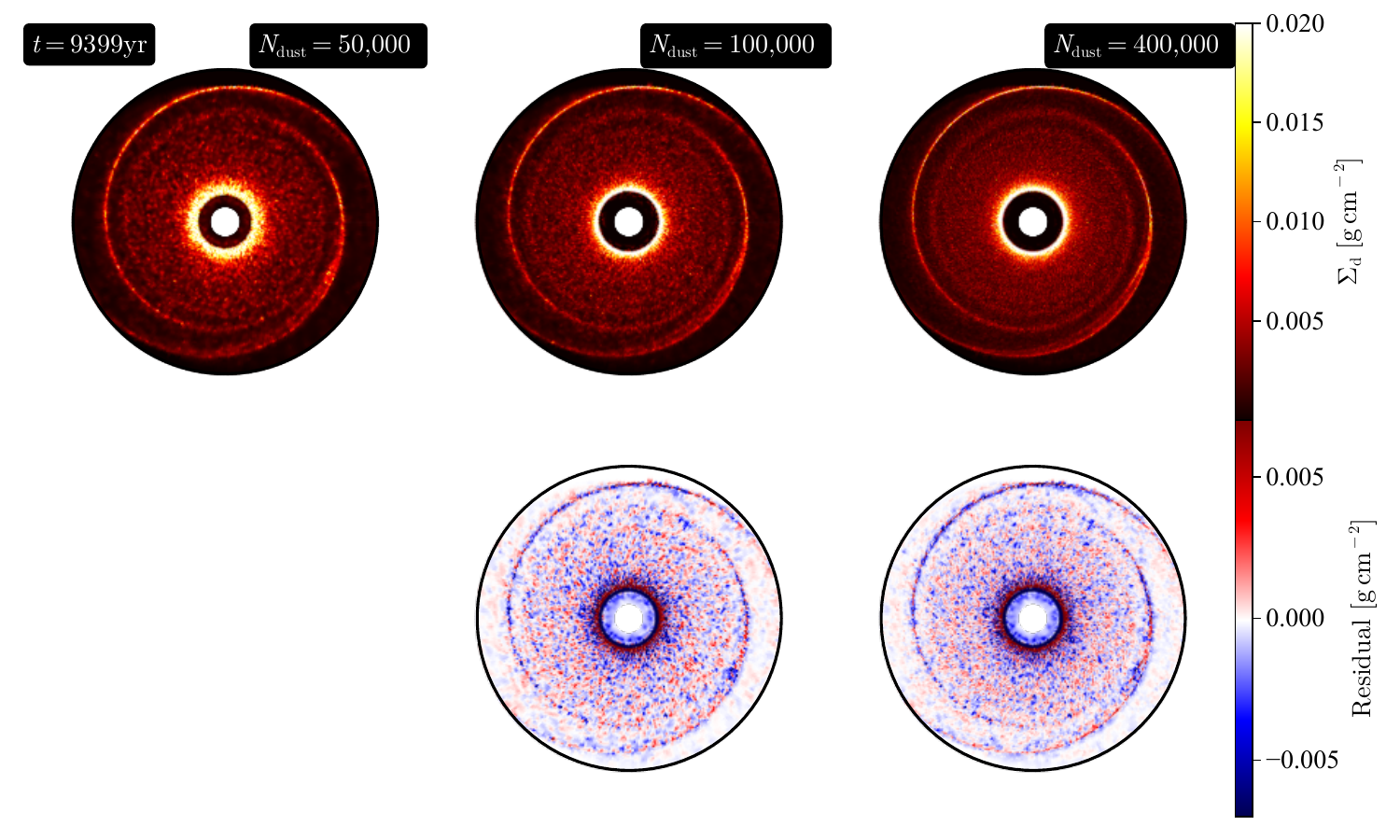}
    \caption{Comparison among three different resolutions. The upper row shows the dust column density, with the three columns representing $N_{\mathrm{dust}} = 5\times 10^4$ (st15co, used in the main text), $N_{\mathrm{dust}} = 10^5$ (st15coHR1), and $N_{\mathrm{dust}} = 4\times 10^5$ (st15coHR2), respectively. The lower row shows the residuals, which are obtained by subtracting the dust column density of st15co from those of the high-resolution simulations}
    \label{fig:ResolutionStudy}
\end{figure*}
To test the effect of numerical resolution on the formation of dust spirals, we performed two high-resolution simulations: st15coHR1 ($N_{\rm dust} = 10^5$) and st15coHR2 ($N_{\rm dust} = 4\times 10^5$). Both simulations were initialized with $10^6$ gaseous Lagrangian particles. Apart from the increased number of dust particles, all other parameters are identical to our fiducial run, st15co (see \tref{table:setuptable}). \fref{fig:ResolutionStudy} compares the resulting dust column densities from these runs. The residuals are obtained by subtracting the dust column density of st15co from those of the high-resolution simulations, given by
\begin{equation}
    \mathrm{Residual} =\Sigma_{\mathrm{d},\mathrm{st15coHR1/2}} - \Sigma_{\mathrm{d},\mathrm{st15co}}
\end{equation}

The residual maps of the higher-resolution simulations show that the dust spiral fails to extend into the inner disc. This indicates that the absence of inner spirals is a physical feature, not a numerical artifact caused by insufficient resolution.
            
%%%%%%%%%%%%%%%%%%%%%%%%%%%%%%%%%%%%%%%%%%%%%%%%%%

% Don't change these lines
\bsp	% typesetting comment
\label{lastpage}
\end{document}